%% file: lattice-3d-skyrm.tex
\pdfoutput=1
\documentclass[twocolumn,aps,superscriptaddress,10pt,nofootinbib,prx]{revtex4-2}
\usepackage[utf8]{inputenc}
\usepackage{amssymb,amsmath,amsfonts,bbm,dsfont,graphicx,hyphenat,bm,xcolor}
\usepackage[english]{babel}
\usepackage[colorlinks=true,linkcolor=blue!60!black,citecolor=blue!60!black, urlcolor=blue!60!black]{hyperref}
\usepackage{newtxtext}
\usepackage{float}
\usepackage[varvw]{newtxmath}
\usepackage[caption=false]{subfig}
\usepackage[percent]{overpic}
\input{commands}

\begin{document}
\title{
Phase diagram of magnetic $S^3$ Skyrmions \\ on three-dimensional lattices and the toroidal antiSkyrmion
}

\author{Niccolò Francini}
\email{niccolo.francini@tu-dresden.de}
\affiliation{Institut f\"ur Theoretische Physik and W\"urzburg-Dresden Cluster of Excellence ct.qmat, TU Dresden, 01062 Dresden, Germany}

\author{Stefano Bolognesi}
\email{stefano.bolognesi@unipi.it}
\affiliation{Dipartimento di Fisica dell’Universit\'a di Pisa and INFN, Largo Pontecorvo 3, I-56127 Pisa, Italy}

\author{Sven Bjarke Gudnason}
\email{gudnason@henu.edu.cn}
\affiliation{Institute of Contemporary Mathematics, School of Mathematics and Statistics, Henan University, Kaifeng, Henan 475004, P.~R.~China}
\affiliation{Department of Physics, Chemistry and Pharmacy, University of Southern Denmark,
Campusvej 55, 5230 Odense M, Denmark}

\author{Roberto Menta}
\email{roberto.menta@sns.it}
\affiliation{NEST, Scuola Normale Superiore, I-56127  Pisa, Italy}

\date{\today}
\begin{abstract}
Magnetic Skyrmions are planar solitons stabilized by the Dzyaloshinskii–Moriya interaction (DMI) and realized in chiral magnets. We study their natural three-dimensional generalization: a sigma model from $ \mathbb{R}^3$ to $S^3$ with a four-component magnetization vector, stabilized by a one-derivative term which is a generalized DMI. We utilize two $\SO(3)$-invariant generalized DMIs discovered recently: an “$\alpha$-term" supporting a spherically symmetric hedgehog Skyrmion and a “$\beta$-term" supporting an axially symmetric Skyrmion that splits into two half-Skyrmions connected by a magnetic string of negative tension, a phenomenon we call “anti-confinement". We derive a cubic-lattice discretization that reproduces both continuum theories at long wavelengths and use Monte Carlo simulations to map the finite-temperature phase diagram. We identify spin-spiral, magnetic-string-lattice, Skyrmion-lattice, and antiSkyrmion-lattice phases, as well as a mixed-topology regime with fractional $S^3$ charges localized at string bends.
We find, for the first time in the literature to the best of our knowledge, a toroidal (anti-)soliton of unit charge.
Our results establish a theoretical and computational framework for three-dimensional topological magnetic textures in systems whose order-parameter manifold is $S^3$.
\end{abstract}

\maketitle

\section{Introduction}
The magnetic Skyrmion is a soliton of two-dimensional magnetic systems, realized experimentally first in chiral magnets and subsequently in a wide range of other materials~\cite{1989JETP...68..101B,1995JETPL..62..247B,Nagaosa2013,doi:10.1126/science.1166767,Yu2010,fert2017magnetic,everschor2018perspective,luo2021Skyrmion,psaroudaki2023Skyrmion}.
Its name derives from the three-dimensional soliton of the Skyrme model, originally proposed as a baryon model 
being a low-energy effective model for QCD
\cite{skyrme1961non,skyrme1962unified}.
In chiral magnetism, a Skyrmion is a spin texture-type soliton whose topological charge counts the winding of the Euclidean plane $\mathbb{R}^2$ onto the target space $S^2$, the space of magnetization vectors~\cite{roessler2006spontaneous,Rossler:2010st,Ezawa:2010uy,Banerjee:2014hna,melcher2014chiral,Rybakov:2018bxt,schroers2019gauged,barton2020magnetic,kuchkin2020magnetic,ross2021Skyrmion,hill2021chiral,schroers2021solvable,Amari:2022boe,Hanada:2023lnm,Bolognesi:2024mjs,Barton-Singer:2024clx, Zheng2026}.
It is stabilized by a single-derivative operator, the Dzyaloshinskii--Moriya interaction (DMI)~\cite{dzyaloshinsky1958thermodynamic,moriya1960anisotropic,moriya1960new}.
Because the underlying theory is chiral, the Skyrmion and the antiSkyrmion are inequivalent, and only one of them is stable for a given sign of the DMI.
This is in contrast to the QCD Skyrmion -- and to its lower-dimensional analog, the baby Skyrmion -- where Lorentz and parity invariance force stabilization by a higher-derivative operator with more derivatives than the spatial dimension.
For magnetic Skyrmions, the lower-derivative DMI term can evade Derrick's theorem~\cite{Derrick:1964ww} provided it contributes negatively to the total energy.

A natural question is whether an analogous construction exists in three spatial dimensions, with a target space supporting a nontrivial $\pi_3$ homotopy group.
The paradigmatic physical example is the B phase of superfluid ${}^3$He, whose order-parameter manifold supports a rich family of three-dimensional topological textures~\cite{Volovik:1977,10.1093/acprof:oso/9780199564842.001.0001}.
A second example is provided by the Skyrme model itself, in which a periodic array of baryon Skyrmions realizes a condensed-matter-like system with a rigid-body order parameter at each lattice site.
More generally, an effective $S^3$ target space arises whenever the local order parameter is a rigid rotor, as occurs for instance in spin-nematic and biaxial-nematic phases of frustrated magnets and spinor condensates~\cite{stamper2013}.
A complementary route to such physics, which we expect to become increasingly relevant, is offered by synthetic dimensions in cold-atom and photonic platforms, where extra target-space directions can be engineered through internal degrees of freedom.
We are not aware of a microscopic realization of the generalized DMI considered below; the present work should therefore be read as a theoretical proposal that anticipates such a realization.

In Ref.~\cite{Gudnason:2024opf} we initiated this higher-dimensional generalization by considering maps from $\mathbb{R}^3$ to the three-sphere $S^3$, retaining the term ``magnetization vector'' for the four-component order parameter despite its more general physical content.
The first task was to classify one-derivative analogs of the DMI, compatible with the largest possible residual symmetry.
We identified two such terms, both invariant under the diagonal subgroup $\SO(3)_{\rm diag}$ of the combined spatial $\SO(3)_{\rm spt}$ and internal $\SO(3)_{\rm int} \subset \SO(4)$ rotations -- the symmetry left unbroken by an effective ``spin--orbit'' coupling.
Up to redefinitions by the broken symmetries, these two terms, which we denote the $\alpha$- and $\beta$-terms, exhaust the possibilities~\cite{Gudnason:2024opf}.
The $\beta$-term vanishes identically on any spherically symmetric Skyrmion Ansatz, while the $\alpha$-term gives a nonzero contribution there.

In the present work, we study both models in detail, supplementing each generalized DMI with a kinetic (Dirichlet) term and a suitable potential.
For the $\beta$-theory we choose a potential that is \emph{not} aligned with the DMI, breaking the residual symmetry to a diagonal $\SO(2)$.
In this setting, the DMI couples nontrivially to the topology already at the level of an axially symmetric Ansatz, and a topologically nontrivial three-dimensional Skyrmion exists provided the DMI is sufficiently strong.
The same model also supports non-topological magnetic strings, whose tension can be tuned to be negative.
When this happens, the three-dimensional Skyrmion splits into two half-Skyrmions connected by a string of negative energy density, a phenomenon we call \emph{anti-confinement}, by analogy with, but opposite to, the confining string of QCD.
In a finite box with Dirichlet boundary conditions, the two half-Skyrmions are localized at the boundary.
For the $\alpha$-theory the spherically symmetric Ansatz already produces a nonzero DMI contribution, and a spherical hedgehog Skyrmion exists when the potential is aligned with the DMI, as analyzed in Ref.~\cite{Gudnason:2024opf}.
We show here that this hedgehog solution is unstable above a critical DMI strength through an anti-confinement mechanism analogous to the one operating in the $\beta$-theory, an instability that is invisible within the hedgehog Ansatz itself.

We then formulate both theories on a cubic lattice such that the corresponding continuum models are recovered at long wavelengths, derive the lattice realizations of the two generalized DMI terms, and perform Monte Carlo simulations of their finite-temperature phase diagrams.
We find a rich phase diagram containing several distinct ordered phases: a spin spiral, a lattice of magnetic strings, a three-dimensional Skyrmion lattice, and an antiSkyrmion lattice.
Interestingly, the antiSkyrmion turns out to be of toroidal shape, which usually happens only for charge-two solitons: our antiSkyrmion is the first -- to the best of our knowledge -- unit-charge soliton with a toroidal shape.
Beyond these, we identify a mixed-topology regime in which fractional $S^3$ charges localize at bends in the magnetic strings; a simple geometric argument predicts the charge per bend to be $1/6$, in agreement with the value $\approx 0.19$ measured on the lattice.
Together, the continuum analysis and the lattice phase diagram establish a framework for three-dimensional topological magnetic textures in systems whose effective order-parameter manifold is $S^3$.

The paper is organized as follows.
In Sec.~\ref{sec:continuumlattice} we develop the continuum theory and analyze the $\alpha$- and $\beta$-models in turn.
In Sec.~\ref{sec:numericallattice} we present the lattice formulations and the Monte Carlo results.
We conclude in Sec.~\ref{sec:conclusion} with a discussion of open questions and candidate physical realizations.
We relegate some details to the appendices: The Luttinger-Tisza method is reviewed in App.~\ref{appendix:LT-method}, the spin-spiral phase is reviewed in App.~\ref{appendix:spin-spiral-continuum-method} and finally, the alignment of string with the SO(3) embedding in the SO(4) group of the magnetization vector is demonstrated in App.~\ref{appendix:alignment_strings}.

\section{Higher-dimensional magnetic Skyrmions}
\label{sec:continuumlattice}
A possible extension of planar magnetic Skyrmions can be realized by increasing the dimension of the base space and target space from two to three  dimensions~\cite{menta2023magnetic, Gudnason:2024opf}. The magnetization field is now a map $\bm{n} : \mathbb{R}^3 \to S^3$ where $\bm{n}=(n^1,n^2,n^3,n^4)$ and $\bm{n}\cdot\bm{n}=1$.
Let us consider the energy functional of the following form:
\begin{eqnarray}
E[\bm{n}] = \sum_{\ell=0,1,2} E_{\ell}[\bm{n}] \ , \qquad
E_{\ell}[\bm{n}]  = \int\d^3x \; \E_{\ell}(\bm{n})\ .
\label{Emagsky}
\end{eqnarray}
The Heisenberg exchange term is
\begin{eqnarray}
\E_2(\bm{n}) = \frac{1}{2} \partial_i\bm{n}\cdot\partial_i\bm{n}
\label{E2cont}
\end{eqnarray}
with $\partial_i$ being spatial derivatives in $\mathbb{R}^3$.
As a potential term $\E_0$, we consider the generalized Zeeman potential:
\begin{eqnarray}
\E_0(\bm{n}) = m^2(1 - \bm{n}\cdot\bm{N})\ ,
\label{potcont}
\end{eqnarray}
that is symmetric around the physical vacuum $\bm{N}$, which is defined by 
\begin{eqnarray}
\lim_{ r \to\infty} \bm{n}= \bm{N} \ .
\label{finite-energy-condition3D}
\end{eqnarray}
The condition in Eq.~\eqref{finite-energy-condition3D} allows for the topological compactification of the base space $\mathbb{R}^3$ into $S^3$, such that the vector field becomes a map $\bm{n}: S^3 \to S^3$. This leads to a nontrivial associated homotopy group $\pi_3(S^3) = \mathbb{Z}\ni Q_{S^3}$, with $Q_{S^3}$ being the topological degree or the so-called ``baryon number'': 
\begin{eqnarray}
 Q_{S^3} 
= \frac{1}{2\pi^2}\int \epsilon_{abcd}n^a\p_1 n^b\p_2 n^c \p_3 n^d\;\d^3x,
\label{eq:topocharge}
\end{eqnarray}
which is the pullback of the normalized volume form on the target space $S^3$ by the vector field $\bm{n}$.
The vacuum can be chosen as $\bm{N}=(0,0,0,1)^{\top}$ and the potential breaks the symmetry from $\Og(4)$ to $\Og(3)$ in the orthogonal space to $\bm{N}$. $E_1$ is the \emph{new} energy term, which is the three-dimensional generalization of the planar DMI.

\subsection{Generalized \texorpdfstring{$\SO(3)$}{SO(3)} invariant DM interactions}\label{sec:genDM}
We review
a possible generalization of from two three dimensions (see Ref.~\cite{Gudnason:2024opf, menta2023magnetic} for details) of the DMI term of Bloch-type,
\begin{eqnarray}
\kappa\epsilon^{iab}\p_in^a n^b, \qquad i=1,2,\qquad a,b=1,2,3,
\label{eq:DM2d_Bloch}
\end{eqnarray}
for an $\SO(4)$ vector. Using  the $\SO(4)$ invariant antisymmetric tensor $\epsilon^{abcd}$, which is totally antisymmetric, with a constant tensor $\Theta^{abi}$, we identify \cite{Gudnason:2024opf}
\begin{eqnarray}
\E^{\rm Bloch}_1 := \kappa \epsilon_{a b c d} \Theta^{abi} \p_i n^c n^d,
\label{eq:DM}
\end{eqnarray}
where $\Theta$ is a mixed tensor having two $\SO(4)$-indices ($a$ and $b$) and one spatial index $i$. We use $\SO(4)$ and $\SO(3)_{\rm spt}$ transformations to
simplify the tensor $\Theta$ as much as possible.
We restrict ourselves to invariant cases under a combined symmetry $\SO(3)_{\rm int}\subset \SO(4)$ and $\SO(3)_{\rm spt}$, which is attained by the tensor
\begin{eqnarray}
\Theta^i = \frac12
\begin{pmatrix}
  0 & \alpha\delta^{i3} & -\alpha\delta^{i2} & \beta\delta^{i1}\\
  -\alpha\delta^{i3} & 0 & \alpha\delta^{i1} & \beta\delta^{i2}\\
  \alpha\delta^{i2} & -\alpha\delta^{i1} & 0 & \beta\delta^{i3}\\
  -\beta\delta^{i1} & -\beta\delta^{i2} & -\beta\delta^{i3} & 0
\end{pmatrix},
\label{eq:Theta_inv_standard}
\end{eqnarray}
with $\alpha,\beta\in\mathbb{R}$. From here on, we will refer to the locked rotations in orbital and internal spin space as $\SO(3)_{\rm diag}$. We see that the invariant tensor only has two degrees of freedom.

One may also consider a simplified version:
\begin{eqnarray}
\Theta^{abi} = \frac12\left(\Gamma^a\delta^{bi} - \Gamma^b\delta^{ai}\right).
\end{eqnarray}
Invariance is only preserved if \cite{Gudnason:2024opf}
\begin{eqnarray}
\Gamma_{\rm inv}^a = -\beta\delta^{a4} 
\label{eq:Gamma_inv}
\end{eqnarray}
in agreement with the $\beta$-part of the invariant tensor in the standard form of Eq.~\eqref{eq:Theta_inv_standard}. 

Another simplified tensor may be constructed by taking the ``Hodge dual'' of the $\Theta$ tensor. To dualize in four dimensions, we extend the index $i$ to $i=1,2,3,4$, but with the derivative $\p_4=0$. In this case, we have \cite{Gudnason:2024opf}
\begin{eqnarray}
\Theta^{abi} = \frac12\epsilon^{abic}\Psi^c,
\label{eq:Theta_Psi}
\end{eqnarray}
where the factor of $1/2$ is introduced for convenience. It is again clear that only a subset of this tensor is invariant under the combined diagonal symmetry $\SO(3)_{\rm diag}$, and it is given by
\begin{eqnarray}
\Psi^a = \alpha\delta^{a4}.
\label{eq:Theta_Psi_inv}
\end{eqnarray}
This corresponds to the $\alpha$-part of the invariant tensor on standard form, see Eq.~\eqref{eq:Theta_inv_standard}.

In two dimensions, the DMI term can equivalently take a different form, known as the DMI of N\'eel type:
\begin{eqnarray}
\kappa(\bm{n}\cdot\nabla n^3 - n^3\nabla\cdot\bm{n}),
\label{eq:DM2d_Neel}
\end{eqnarray}
with $\nabla=(\p_1,\p_2,0)$ and $\bm{n}=(n^1,n^2,n^3)$, which is due to the Rashba spin-orbit coupling (SOC), whereas the Bloch-type DMI in Eq.~\eqref{eq:DM2d_Bloch} corresponds to the Dresselhaus SOC.
Although this DMI term looks quite different from the Bloch type in Eq.~\eqref{eq:DM2d_Bloch}, the N\'eel-type DMI is simply obtained from the Bloch type DMI by performing an {$\SO(2)_{\rm int} \subset \SO(3)_{\rm int}$} rotation of the magnetization vector $(n^1,n^2,n^3)\mapsto(n^2,-n^1,n^3)$ and keeping the derivative vector fixed.
However, the generalization of the above two-dimensional DMI term to a higher-dimensional DMI term is straightforward and somewhat different~\footnote{Sometimes the opposite sign of the N\'eel-type DMI is used, which can be reached from the Bloch-type DMI simply by rotating the opposite way in the 1-2 plane: $(n^1,n^2,n^3)\mapsto(-n^2,n^1,n^3)$, or by flipping the sign of $\kappa$.},
\begin{equation}
\E^{\text{N\'eel}}_1 = \kappa(n^4\nabla\cdot\bm{n} - \bm{n}\cdot\nabla n^4)
=\frac12\kappa\epsilon_{abcd}\epsilon^{abi4}\p_in^c n^d,
\label{eq:DM_Neel}
\end{equation}
where now $\nabla=(\p_1,\p_2,\p_3,0)$ and $\bm{n}=(n^1,n^2,n^3,n^4)$. 
So the N\'eel type DMI interaction is just a different way to write the $\alpha$-type tensor, as can be seen by comparison with Eqs.~\eqref{eq:Theta_Psi} and \eqref{eq:Theta_Psi_inv}. This way of writing the generalized DMI will be particularly useful when we perform the discretized version of the $\alpha$-type later in Sec.~\ref{sec:lattice}.
%

\subsection{Skyrmion Ansatz}
\label{subsec:Skyrmion-ansatz}
The hedgehog Ansatz with an azimuthal rotation by $\delta$ is given by the map
\begin{equation}
\bm{n} = \begin{pmatrix}
	  \sin\chi\sin\theta\cos(\phi+\delta)\\
	  \sin\chi\sin\theta\sin(\phi+\delta)\\
	  \sin\chi\cos\theta\\
	  \cos\chi
	\end{pmatrix},  \quad
	\chi,\theta\in[0,\pi], \quad
	\phi\in[0,2\pi),
	\label{eq:hedgehog}
\end{equation}
with $(r,\theta,\phi)$ being the normal spherical coordinates and $\chi(r)$ the radial profile function.
The {$\SO(3)_{\rm diag}$}-invariant DMI term of Eq.~\eqref{eq:Theta_inv_standard}, evaluated on the hedgehog Ansatz of Eq.~\eqref{eq:hedgehog} reduces to
\begin{align}
  \E_1^{\rm Bloch} &= 
  \kappa\alpha\cos\delta\left(\sin^2\theta\chi' + \frac{(1+\cos^2\theta)\sin2\chi}{2r}\right)
  \non  &\phantom{=\ } + \kappa\alpha\left(\cos^2\theta\chi' + \frac{\sin^2\theta\sin2\chi}{2r}\right)\non
  &\phantom{=\ }
  + 2\kappa\beta\sin\delta\,\frac{\cos\theta\sin^2\chi}{r}.
  \label{eq:block-type-DMI-spheric}
\end{align}
Integrating Eq.~\eqref{eq:block-type-DMI-spheric} over $\theta$, we obtain
\begin{eqnarray}
\int \E_1^{\rm Bloch}\sin\theta\d\theta
= \frac{2\kappa\alpha}{3}(1+2\cos\delta)\left(\chi' + \frac{\sin2\chi}{r}\right).
\end{eqnarray}
The $\beta$-part of the term vanishes upon integration of $\d\theta$. The $\alpha$-term becomes spherically symmetric when $\delta=0$:
\begin{eqnarray}
\E_1^{\rm Bloch} = \kappa\alpha\left(\chi' + \frac{\sin2\chi}{r}\right).
\end{eqnarray}
Since the DMI term contributes as a negative energy, maximizing its prefactor corresponds to minimizing the energy, which is quickly seen to occur at $\delta=0$, the symmetric point.
If we consider the generalized N\'eel-type DM term in Eq.~\eqref{eq:DM_Neel}, we obtain the $\alpha$-part of the Bloch-type DMI (see Eq.~\eqref{eq:block-type-DMI-spheric}) under the hedgehog Ansatz, as expected.
%

\subsection{A theory with the \texorpdfstring{$\beta$}{beta}-term}\label{sec:beta_continuum_theory}
Consider the magnetic Skyrmion discussed above with the $\beta$-term only in the DMI (see Eq.~\eqref{eq:Theta_inv_standard}) and 
\begin{eqnarray}
\bm{N}=\begin{pmatrix}
	0 ,	0 , 0, 1
\end{pmatrix}^{\top}, \quad  \qquad 
\bm{\Gamma}=\begin{pmatrix}
	0 ,	0 ,		\sin \gamma,\cos \gamma
\end{pmatrix}^{\top}\ .
\label{ng}
\end{eqnarray}
where the first $\SO(4)$-transformation  rotates $\bm{N}$ in the $\bm{e}_4$-direction, while the second rotation of the unbroken $\SO(3)$ can rotate the remaining part of $\bm{\Gamma}$ in the $\bm{e}_3$-direction.
The angle $\gamma$ is an intrinsic parameter of the model, and $\bm{N}$ will correspond to the vacuum field of the theory. 
In this model, for the aligned case $\gamma=0$, as discussed before, any spherical symmetric Skyrmion does not feel the DMI term. Therefore we consider here the non-aligned case $\gamma \neq 0$, and we will specialize to the simplest case $\gamma=\frac{\pi}{2}$ in the following.
However, deformed solitons may feel the presence of the $\beta$-part of the DMI term, which in principle could stabilize a Skyrmion against collapse.

The symmetry group of the $\SO(4)_{\rm int}$ sigma model undergoes a breakdown to $\SO(3)_{\rm int}$ due to the magnetic potential that breaks symmetry, in particular by the vacuum field $\bm{N}$ in $E_0$. Consequently, the combination $E_2 + E_0$ encompasses spatial and internal rotations, forming $\SO(3)_{\rm spt}\times \SO(3)_{\rm int}$.
Additionally, there exist two internal parities as a discrete group, denoted as $P_{\rm spt}\times P_{\rm int}$.
The subsequent step involves considering the DMI term. This term effectively couples space rotations with internal rotations. When $\bm{\Gamma}\propto\bm{N}$, i.e., $\gamma=0$, the interaction term remains invariant under simultaneous transformations in both spaces, leading to $\SO(3)_{\rm diag}$ for the continuous group and $P_{\rm diag}$ for the discrete group. Here we denote with $P_{\rm diag}$ the simultaneous parity operation in orbital and spin spaces.
However, if $\gamma$ is not equal to $0$ or $\pi$, only $\SO(2)_{\rm diag}$ remains unbroken. We thus expect the Skyrmion to be at most axially symmetric. 

The target space $S^3$ defined by the unitary length $\bm{n}\cdot \bm{n}=1$, is parameterized by three Euler angles $(\chi,\vartheta,\varphi)$:
\begin{equation}\label{3sphericalcoord}
	\bm{n}=\begin{pmatrix}
		\sin\chi \sin\vartheta \cos\varphi \\
		\sin\chi \sin\vartheta \sin\varphi \\
		\sin\chi \cos\vartheta \\
		\cos\chi \\
	\end{pmatrix},\quad
	\chi,\vartheta \in [0,\pi]\ ,\quad
	\varphi \in [0,2\pi) \ .
\end{equation}
The metric of the target sphere $S^3$ in these coordinates is  $
	\d s^2 = \d\chi^2 + \sin^2\chi\big(\d\vartheta^2 + \sin^2\vartheta\d\varphi^2\big) \ .
$
The standard terms in the Skyrme model are
\begin{align}
	\E_0(\bm{n}) &=  m^2(1-  \cos{\chi}  )\ , \non
	\E_2(\bm{n}) &= \frac{1}{2}\left[(\partial_i \chi)^2
	+ \sin^2\chi (\partial_i \vartheta)^2
	+ \sin^2\chi  \sin^2\vartheta (\partial_i \varphi)^2\right], 
\end{align}
whereas the generalized DM term for the case $\gamma = \pi/2$ is 
\begin{align}
	\ \E_1(\bm{n}) =  
	\; -\kappa  \,
	\det\begin{pmatrix}
		\sin\chi \sin\vartheta \cos\varphi &
		\sin\chi \sin\vartheta \sin\varphi &
		\cos\chi \\
		\partial_2&\partial_3&\partial_1\\
		\sin\chi \sin\vartheta \cos\varphi &
		\sin\chi \sin\vartheta \sin\varphi &
		\cos\chi  
	\end{pmatrix}\,,
\end{align}
which is like the standard Bloch-type DMI in every $S^2$ sphere at $n_4$ constant.

\begin{figure}[h]
	\centering
	\includegraphics[width=\columnwidth]{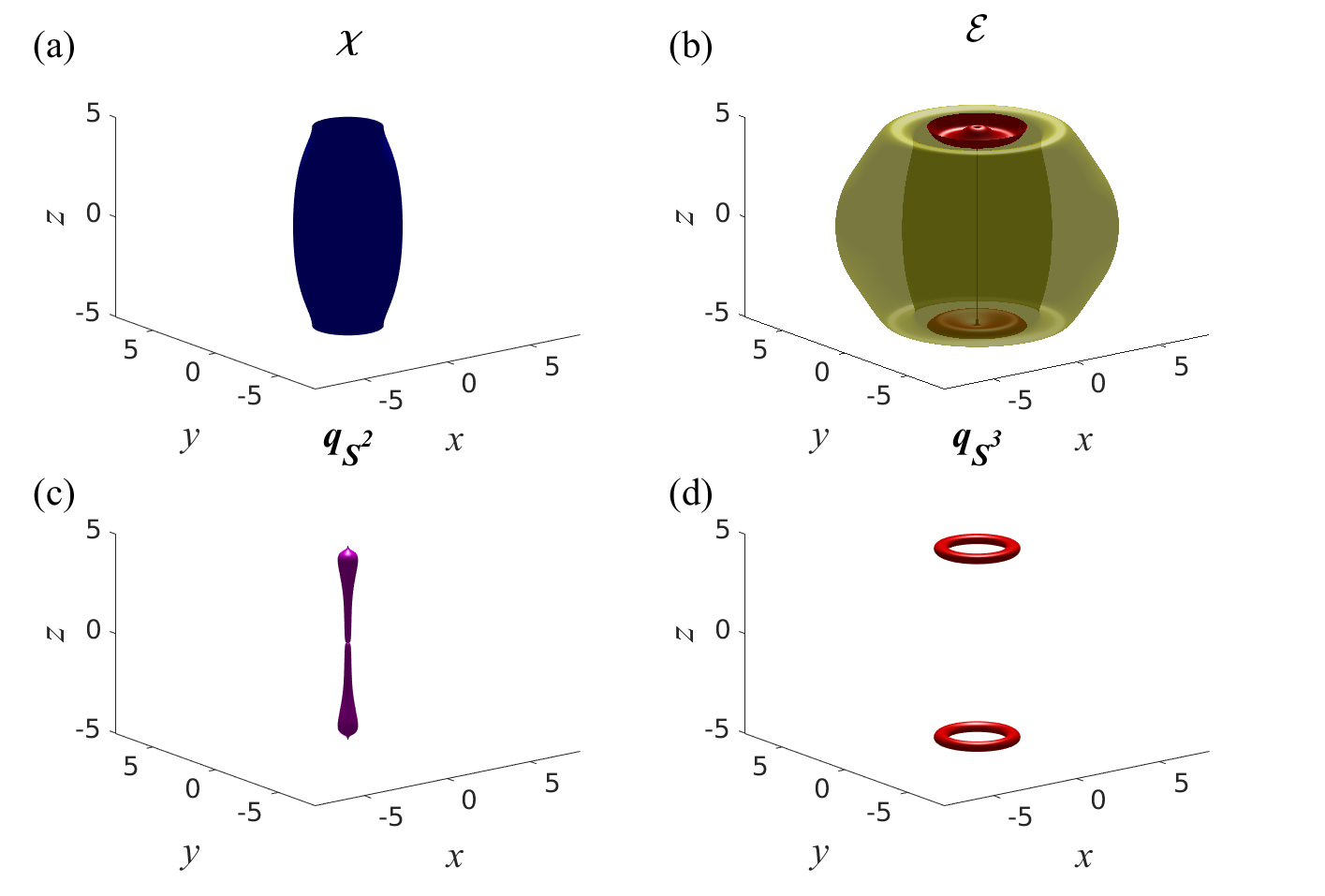} 
	\caption{Numerical solution of the axially symmetric baryon with Dirichlet boundary conditions. The equatorial magnetic string pushes two halves of the Skyrmion to the boundaries of the box.
    (a) The profile function $\chi$ of Eq.~\eqref{eq:axial_Ansatz_1}, (b) The total energy density with the red isosurface showing the positive region and the yellow isosurface showing the negative region, (c) the $q_{S^2}$ topological charge density demonstrating the presence of the string, and (d) the $q_{S^3}$ topological charge density showing the two half-Skyrmions' charge at the end of the string.
    }
	\label{cigarsolution}
\end{figure}

Let us now instead consider the  axially symmetric Ansatz,
\begin{eqnarray}
\bm{n} =
\begin{pmatrix}
	\sin(\chi)\sin(\vartheta)\cos(\varphi)\\
	\sin(\chi)\sin(\vartheta)\sin(\varphi)\\
	\sin(\chi)\cos(\vartheta)\\
	\cos(\chi)
\end{pmatrix},
\label{eq:axial_Ansatz_1}
\end{eqnarray}
with $\chi=\chi(r,\theta)$ and $\vartheta=\vartheta(r,\theta)$  and $\varphi = \phi - \pi/2$.  
The Skyrmion may have a different orientation $\SO(3)$, but the one in Eq.~\eqref{eq:axial_Ansatz_2} should be the preferred one.
We can simplify the reduced energy to 
\begin{align}
	E = 2\pi\int\;\bigg[&
	\frac12\chi_r^2
	+\frac{1}{2r^2}\chi_\theta^2
	+\frac12\sin^2(\chi)\vartheta_r^2
	\non  & +\frac{1}{2r^2}\sin^2(\chi)\vartheta_\theta^2
	+\frac{1}{2r^2\sin^2\theta}\sin^2\chi\sin^2\vartheta\non
	& +\kappa \bigg( \frac{1}{r \sin \theta} \sin \chi \cos \chi  \sin \vartheta \non &  \qquad + \frac{\cos \theta}{r }  \big( \vartheta_{\theta}\sin \chi \cos \chi  \cos \vartheta+   \chi_{\theta} \sin \vartheta \big) \non
	& \qquad  + \sin \theta \big( \vartheta_r\sin \chi \cos \chi \cos \vartheta+  \chi _r  \sin \vartheta \big)\bigg) \non
	&
	+ m^2(1-\cos\chi)
	\bigg]r^2\sin\theta\;\d r\d\theta,
	\label{eq:En_reduced_simplified}
\end{align}
where the subscripts refer to the derivatives, for instance $\chi_r := \partial_r \chi$.
Suitable boundary conditions are
\begin{align}
\chi(0,\theta)&=\pi,\qquad&
\chi(\infty,\theta)&=0, \non
\p_\theta\chi(r,0)&=0,\qquad&
\p_\theta\chi(r,\pi)&=0,
\end{align}
and
\begin{align}
\p_r\vartheta(0,\theta)&=0,\qquad&
\p_r\vartheta(\infty,\theta)&=0, \non 
\vartheta(r,0)&=0,\qquad&
\vartheta(r,\pi)&=\pi,
\end{align}
for which $\chi_r$ is generally negative and hence the stabilizing
sign for $\kappa$ is
\begin{eqnarray}
\kappa>0 \ .
\end{eqnarray}
The DMI does not contain the derivative with respect to $z$ so we do not expect to ever obtain a stable Skyrmion stabilized in all three directions.
If we consider the $\theta = \pi/2$ plane, we have that
\begin{eqnarray}
\bm{n} =
\begin{pmatrix}
	 \sin(\chi)\cos(\varphi)\\
	 \sin(\chi)\sin(\varphi)\\
	 0\\
	\cos{(\chi)}
\end{pmatrix},
\label{eq:axial_Ansatz_2}
\end{eqnarray}
with $\chi=\chi(r,\pi/2)$, $\vartheta=\vartheta(r,\pi/2)$ and $\varphi = \phi - \pi/2$, which is exactly the magnetic string. For sufficiently large $\kappa$, the magnetic string acquires negative tension.
A Skyrmion with unitary topological number is thus expected to minimize the energy by creating such a string. This effect is a sort of opposite of confinement as the negative tension string splits the Skyrmion into two halves and sends them as far as possible from one another~\footnote{It is not exactly ``deconfinement'', as there is a negative energy density localized on the string.}. Thus, we may call it ``anti-confinement''. To observe the Skyrimon, we must stabilize the IR divergence. One way is to solve the equations of motion on a finite box with a finite-difference approximation to the derivatives and Dirichlet boundary conditions, as in Fig.~\ref{cigarsolution}. The solution looks like a cigar with the two ends corresponding to the two anti-confined half-Skyrmions, with the topological charge being localized close to the boundary of the space. The negative energy density is, instead, localized in the whole magnetic string. 

The $\beta$-term model with generic $\gamma$ has a magnetic string as long as $\gamma>0$. Due to the parametrization in Eq.~\eqref{ng}, the string is always in the $z$-direction.
For $\gamma = \pi/2$ we have a special case, and the magnetic string is simply embedded in the $(n_3=0)$-$S^2$ sphere and splits the $S^3$ sphere, and the Skyrmion, exactly in two equal parts. For generic $\gamma$ we expect that the magnetic string will still exist and again oriented in the $z$-direction, but the embedding could be more complicated and the Skyrmion will split in two asymmetric parts. Finally $\gamma=0$ should be a trivial case with no magnetic string and no Skyrmion.

It is worth noticing that the magnetic string is not exactly topological as the equatorial $n_3=0$ does not give the $\pi_2$ topological number when considered embedded in the $S^3$. We may call it {\it semi-topological}, as it arises from a combination of energetic and topological arguments. The negative DMI makes the equatorial $S^2$ the most suitable region in the target space; within this $S^2$ we can have the standard topology of the magnetic string. 
It should be energetically stable when $|\kappa|$ is large enough to make the tension negative. So, there exists a $\kappa_{\rm cr}$ corresponding to tensionless case of the string. For $\kappa<\kappa_{\rm cr}$ there is no string (unlike the two-dimensional case where it is topological) and also no Skyrmion.
%
\subsection{A theory with the \texorpdfstring{$\alpha$}{alpha}-term, or N\'eel term}\label{sec:alpha_continuum_theory}
%
%
%
With the $\alpha$-terms of the Bloch-type DMI, or equivalently the N\'eel term, the spherically symmetric Skyrmion is described by Eq.~\eqref{eq:hedgehog}, and the minimum is reached for $\delta=0$.
The total energy ($E=\int\E\d^3x$) of the higher-dimensional magnetic
Skyrme model under the hedgehog Ansatz in Eq.~\eqref{eq:hedgehog} is thus
given by 
\begin{eqnarray}
E = 4\pi\int\bigg[
\frac12r^2(\chi')^2
+\sin^2\chi
+\kappa(r^2\chi' + r\sin2\chi) \nn \\
\mathop+m^2 r^2(1-\cos\chi)
\bigg] \d r,
\end{eqnarray}
which leads to the following Euler-Lagrange equation
\begin{eqnarray}
\chi'' + \frac2r\chi' - \frac{\sin2\chi}{r^2}
+\frac{4\kappa\sin^2\chi}{r}
- m^2\sin\chi = 0.  
\label{eq:eom_chi}
\end{eqnarray}
Using the Derrick's scaling argument,  $r\to\lambda r$,
\begin{eqnarray}
E_\lambda = \lambda E_2 + \lambda^2 E_1 + \lambda^3 E_0.
\label{eq:Elambda}
\end{eqnarray}
Stability can be attained if $E_1$ is negative. 
By identifying the characteristic radius of
the soliton as $R$ and approximating the integrals of the energy
functional by constants as
\begin{eqnarray}
\lambda E_2 = c_2 R, \quad
\lambda^2 E_1 = -c_1 \kappa R^2, \quad
\lambda^3 E_0 = c_0 m^2 R^3,
\end{eqnarray}
we can write the approximate energy functional as a function of the
soliton size, $R$:
\begin{eqnarray}
E(R) = c_2 R - c_1\kappa R^2 + c_0 m^2 R^3.
\label{eq:ER}
\end{eqnarray}
The virial law is now given by
\begin{eqnarray}
E'(R)  = 0.
\end{eqnarray}
In contradistinction to the two-dimensional magnetic Skyrmion case, in three
dimensions there are two solutions to the virial equation:
\begin{equation}
R_\pm = \frac{\tilde\kappa\pm\sqrt{\tilde\kappa^2-3}}{3}R_0,\quad
\label{eq:Rpm_magskyrm_sphaleron}
\tilde\kappa = \frac{\kappa c_1}{\sqrt{c_0c_2}m}, \quad R_0=\sqrt{\frac{c_2} {c_0}} \frac{1}{m}.
\end{equation}
Clearly, there are no solutions if $\tilde\kappa<\sqrt{3}$, in which
case the DMI term is not strong enough to stabilize a soliton solution.
However, as soon as $\tilde\kappa$ is big enough, two solutions exist:
the large one is the 3D magnetic Skyrmion and we call the small one a magnetic sphaleron.
%
%

\subsection{Lattice realization of the models}
\label{sec:lattice}
%
In the three-dimensional case, the vector field $\bm{n} \in S^3$, is  a four-dimensional vector $\bm{n} = (n^1,n^2,n^3,n^4)$ that satisfies the constraint $|\bm{n}|^2=1$. The lattice Hamiltonian for the Heisenberg term is 
\begin{equation}\label{Heisenberg_discrete}
	\mathcal{H}_{\rm H} = - J \sum_{I, \mu} \bm{n}^{(I)} \cdot \bm{n}^{(I + a \hat{\bm{e}}_{\mu})} \ ,
\end{equation}
where $a \hat{\bm{e}}_{\mu}$ is the vector connecting the $I$-lattice site with its neighboring sites $I + a \hat{\bm{e}}_{\mu}$, on a cubic ($\mu=x,y,z$) lattice. Assuming a small lattice constant $a$, one
can perform a Taylor expansion on $\bm{n}^{(I + a \hat{\bm{e}}_{\mu})}$, i.e.,
\begin{equation}
	\bm{n}^{(I + a \hat{\bm{e}}_{\mu})} = \bm{n}^{(I)} + a \partial_{\mu} \bm{n}^{(I)} + \frac{a^2}{2} \partial^2_{\mu} \bm{n}^{(I)} + \cdots
\end{equation}
and insert it into Eq.~\eqref{Heisenberg_discrete}. The first term $\bm{n}^{(I)} \cdot \bm{n}^{(I)}$ is equal to unity and can be discarded as a constant term. The second term, proportional to $\bm{n}^{(I)}\cdot\partial_{\mu}\bm{n}^{(I)}$, vanishes due to the nonlinear constraint $\bm{n}\cdot\bm{n}=1$.
However, the third term gives the following:
\begin{equation}
    \label{eq:heisenberg-lattice}
	\mathcal{H}_{\rm H} \simeq - \dfrac{1}{2}Ja^2 \sum_I \bm{n}^{(I)} \cdot \partial^2_{\mu} \bm{n}^{(I)} \ .
\end{equation}
In the continuum limit $a \to 0$, and after an integration by parts, Eq.~\eqref{eq:heisenberg-lattice} leads to the Dirichlet term 
\begin{equation}
	\label{eq:Gamma-DM}
	E_2[\bm{n}] = \dfrac{1}{2} J a^{-1} \int \d^3x \  \partial_{\mu} \bm{n} \cdot \partial_{\mu} \bm{n} \ , \qquad \bm{n}\cdot\bm{n}=1 \ .
\end{equation}
The continuum limit is obtained by sending $a \to 0$ while keeping $J a^{-1}$ fixed. $J a^{-1}$ is the squared radius of the $S^3$ sphere in the continuum, which we fixed equal to $1$ in Eq.~\eqref{E2cont}.

A possible candidate for the discrete lattice DM Hamiltonian is 
\begin{equation}
    \label{axial_DM_discrete}
	\mathcal{H}_{\rm DM} = \sum_{I,\mu}\epsilon^{ijk\ell} D_{\mu,i} \Gamma_j n^{(I)}_k n^{(I + a\hat{\bm{e}}_{\mu})}_{\ell}\ ,
\end{equation}
where $n^{(I)}_k$ indicates the spin versor $\bm{n}$ of component $k$  at the site $I$. The DM four-dimensional vector can be expressed as $D_{\mu,i} = D (\hat{\bm{e}}_{\mu})_{i}$ and for the cubic lattice is $D_{\mu,i} = D \delta_{\mu i }$.   $\bm{\Gamma}$ is a constant vector that is necessary to define this type of interaction. 
Upon Taylor expansion for $a\ll 1$, the first derivative of $\bm{n}^{(I  + a \hat{\bm{e}}_{\mu})}$ gives
\begin{eqnarray}
	\mathcal{H}_{\rm DM} &\simeq & Da \sum_{I ,\mu} \epsilon^{ijk\ell} \hat{\bm{e}}_{\mu,i} \Gamma_j n^{(I)}_{k} \partial_{\mu} n^{(I)}_{\ell} \nn \\ &=& Da \sum_I \epsilon^{ijk\ell} \Gamma_i n^{(I)}_{j} \partial_k n^{(I)}_{\ell} \ .
\end{eqnarray}
 The continuum limit leads to
\begin{equation}
	E_1[\bm{n}] = Da^{-2} \int d^3x \ \epsilon^{ijk\ell} \Gamma_i n_{j} \partial_k n_{\ell} \ ,
\end{equation}
which corresponds to a possible generalization of the DM term of Bloch-type, that is, the theory with the $\beta$-term in Sec.~\ref{sec:beta_continuum_theory}.

An alternative DMI can be constructed as follows:
\begin{widetext}
\begin{align}
\label{Neeldiscrete}
\mathcal{H}_{\rm DM,sph} &= \sum_{I,\mu} \left[(\hat{\bm{e}}_4 \cdot \bm{n}_I) (\boldsymbol{D}_\mu\cdot\bm{n}_{I+a \hat{\bm{e}}_\mu})
- (\hat{\bm{e}}_\mu\cdot\bm{n}_I)(\boldsymbol{D}_4\cdot\bm{n}_{I+a\hat{\bm{e}}_\mu}) \right]\non
&=\sum_{I,\mu} \left[n_I^4 \boldsymbol{D}_\mu\cdot(\bm{n}_I+a\p_\mu\bm{n}_I) 
- n_I^\mu\boldsymbol{D}_4\cdot(\bm{n}_I+a\p_\mu\bm{n}_I) \right]\non
&=a\sum_{I,\mu}\left[n_I^4 \boldsymbol{D}_\mu\cdot \p_\mu\bm{n}_I
-n_I^\mu\boldsymbol{D}_4\cdot \p_\mu\bm{n}_I\right].
\end{align}
\end{widetext}
Choosing $\boldsymbol{D}_\mu=D\bm{e}_\mu$, we get
\begin{eqnarray}
\label{eq:lattice-spherical-DM}
\mathcal{H}_{\rm DM}=
aD\sum_{I,\mu}\left[n_I^4 \p_\mu n_I^\mu-n_I^\mu\p_\mu n_I^4\right].
\end{eqnarray}
Taking the continuum limit, the spherical DMI term becomes
\begin{eqnarray}
\label{eq:DMI_sph}
E_{1}[\bm{n}]=D a^{-2}\int d^3x\;
\left[n^4\nabla\cdot\bm{n}-\bm{n}\cdot\nabla n^4\right].
\end{eqnarray}
This construction of the DMI has a manifest maximal symmetry $\SO(3)$ and is the  N\'eel-type DMI or theory with the $\alpha$-term, see Sec.~\ref{sec:alpha_continuum_theory}.

In addition, we add a Zeeman term
\begin{equation}
\label{Zeeman_discrete}
	\mathcal{H}_{\mathrm{Z}}=-\bm{B}\cdot \sum_{I} \bm{n}_I,
\end{equation}
which, in the continuum limit, up to an irrelevant constant, becomes Eq.~\eqref{potcont} with
\begin{eqnarray}
m^2 = \frac{B}{a^3}\;,  \qquad \bm{N} = \frac{\bm{B}}{B}\;.
\end{eqnarray} 
%
\section{Lattice results}
\label{sec:numericallattice}
In this Section, the numerical results for the lattice models are reported. The finite temperature phase diagram is obtained by means of Monte Carlo simulation. We employed local Metropolis updates~\cite{metropolis1953} combined with an overrelaxation algorithm~\cite{creutz1987} and parallel tempering to mitigate freezing effects~\cite{marinari92,hukushima96,hukushima99} at low temperatures. 
The classical zero-temperature ground state is obtained with an iterative minimization algorithm, with the starting spin configuration given by the lowest temperature replica from the Monte Carlo simulations~\cite{janssen2016heisenbergkitaev}. We considered a cubic lattice of $L\times L\times L$ volume with boundary conditions depending on the model. After an initial thermalization time, we accumulated an order of $\mathcal{O}(10^5)$ measures of our observables for each point in the parameter space. The simulations are carried out on the high-performance computer Barnard at the NHR Center of TU Dresden~\cite{nhr-alliance}.
%
\subsection{Observables}   %
In our simulations, we measure several thermodynamic observables as averages from their Monte Carlo histories. In the following, the notation $\langle O\rangle$ indicates the Monte Carlo average for the quantity $O$. Two observables are particularly useful in the understanding of the lattice phase diagram.
\paragraph*{\texorpdfstring{$S^3$}{S3} topological charge.}
From the definition of the $\pi_3(S^3)$ topological charge in the continuum, see Eq.~\eqref{eq:topocharge}, it is possible to obtain a lattice version:
\begin{equation}
    \label{eq:S3-topological-charge}
    Q_{S^3} = \left\langle \frac{1}{2\pi^2}\sum_{I}\sum_{\alpha\beta\mu\nu} \epsilon^{\alpha\beta\mu\nu} S_\alpha^{(I)} S_\beta^{(I+\hat{\bm{e}}_1)} S_\mu^{(I+\hat{\bm{e}}_2)}S_\nu^{(I+\hat{\bm{e}}_3)}\right\rangle,
\end{equation}
where the index $I$ refers to the lattice site, while the greek indices count the spin components.
\paragraph*{\texorpdfstring{$S^2$}{S2} topological charge.}
In the axial case in Eq.~\eqref{axial_DM_discrete}, the choice of $\boldsymbol{\Gamma}=(0,0,1,0)$ simplifies the problem. In the bulk, the third spin component is suppressed to zero, and we can identify effective three-component spins. Hence, we can also define an effective $\pi_2(S^2)$ topological charge, accounting for the winding of the effective three-component magnetization vector. 
For a magnetic moment winding in the $xy$ plane, we have a $\pi_2(S^2)$ topological charge that in the lattice version reads
\begin{equation}
    \label{eq:S2-topological-charge}
    Q_{S^2} = \left\langle \frac{1}{4\pi}\sum_{I} \tilde{\bm{n}}_I \cdot \left(\tilde{\bm{n}}_{I+\hat{\bm{e}}_1} \times \tilde{\bm{n}}_{I+\hat{\bm{e}}_2} \right) \right\rangle,
\end{equation}
where the effective spins are $\tilde{\bm{n}}_I=(n_I^1,n_I^2,n_I^4)$. Strictly speaking, the quantity in Eq.~\eqref{eq:S2-topological-charge} is not really topological, as the embedded $S^2$ manifold with $n^3=0$ is favored by the negative tension of the strings, but it is not enforced by any constraint. 
\paragraph*{Densities of charges.}
The zero-temperature configuration is reached through iterative minimization from the lowest-temperature replica of the Monte Carlo simulation. Once the ground state is obtained, it is convenient to measure the $\pi_2(S^2)$ and $\pi_3(S^3)$ topological charges per site:
\begin{equation}
\label{eq:charge-and-widning-densities}
    \begin{split}
        q_{S^3}^{(I)}&=\frac{1}{2\pi^2}\sum_{\alpha\beta\mu\nu} \epsilon^{\alpha\beta\mu\nu} n_\alpha^{(I)} n_\beta^{(I+\hat{\bm{e}}_1)} n_\mu^{(I+\hat{\bm{e}}_2)} n_\nu^{(I+\hat{\bm{e}}_3)}\,, \\
        q_{S^2}^{(I)}&=\frac{1}{4\pi} \tilde{\bm{n}}_I \cdot \left(\tilde{\bm{n}}_{I+\hat{\bm{e}}_1} \times \tilde{\bm{n}}_{I+\hat{\bm{e}}_2}\right)\,.
    \end{split}
\end{equation}
These quantities are local and bring additional information on the distribution of the $\pi_2(S^2)$ and $\pi_3(S^3)$ topological charges on the lattice, allowing the identification of strings and Skyrmions.
%
\subsection{Lattice theory with the \texorpdfstring{$\beta$}{beta}-term}   %
\label{subsec:axial-DM-numerics}           %
%
The lattice model with the DMI $\beta$-term consists of two nearest-neighbor terms: a ferromagnetic Heisenberg contribution and an axially symmetric Dzyaloshinskii–Moriya interaction. In addition, a Zeeman term is present, with an external field $\bm{B}$ fixing the vacuum direction. Putting together Eqs.~\eqref{Heisenberg_discrete}-\eqref{axial_DM_discrete}-\eqref{Zeeman_discrete}, we obtain
\begin{align}
     \mathcal{H} = &- J \sum_{I, \mu} \bm{n}^{(I)} \cdot \bm{n}^{(I + a \hat{\bm{e}}_{\mu})} + \sum_{I,\mu}\epsilon^{ijkl}D_{i}^{(I + a \hat{\bm{e}}_{\mu})} \Gamma_{j}n_k^{(I)}n_l^{(I + a \hat{\bm{e}}_{\mu})} \nonumber \\
     &-\bm{B}\cdot\sum_{I}\bm{n}^{(I)}\,.
\label{eq:axial-lattice-model}
\end{align}
The DM vector is $D_i^{(I + a \hat{\bm{e}}_{\mu})}=(\hat{\bm{e}}_{\mu})_i$, while $\bm{\Gamma}=(0,0,\sin{\gamma},\cos{\gamma})$. In the following, we consider the case discussed in Sec.~\ref{sec:beta_continuum_theory} with $\gamma=\pi/2$, giving $\bm{\Gamma}=(0,0,1,0)$. The vacuum direction is fixed by the Zeeman field $\bm{B}=B(0,0,0,1)$. The choice for the couplings $J=1$ and $D=\sqrt{2}$, i.e.~$\kappa=D/J=\sqrt{2}$, is motivated by analytical Luttinger-Tisza results for $\bm{B}=0$ in periodic boundary conditions, see App.~\ref{subapp:axialDMI-LT}. However, in the numerical simulations, we considered fixed twisted boundary conditions with 
\begin{equation}
\begin{split}
    \bm{n}^{(I)} &= (0,0,-1,0) \quad \mathrm{for}\,\, I\in \begin{cases}
        (0,y,z)\,, \\
        (x,0,z)\,, \\
        (x,y,0)\,,
    \end{cases} \\
    \bm{n}^{(I)} &= (0,0,1,0) \quad \mathrm{for}\,\, I\in \begin{cases}
        (L-1,y,z)\,, \\
        (x,L-1,z)\,, \\
        (x,y,L-1)\,.
    \end{cases}
\end{split}
\end{equation}
The unusual choice of the boundary conditions is motivated by the continuum theory, see Sec.~\ref{sec:beta_continuum_theory}. Once a string is formed, it carries a half-Skyrmion at its extrema. Periodic boundary conditions are not sufficient to capture this effect, since they would return a closed string with ring shape, losing the Skyrmion at the extrema.
The string solution only requires three out of the four components of the magnetization vector, $\bm n$, and hence a string can be forced to end on the vacuum boundary condition $\bm n=(0,0,0,1)$ without activating one of the components, say $n_3$: such a solution has no $\pi_3(S^3)$ topological charge.
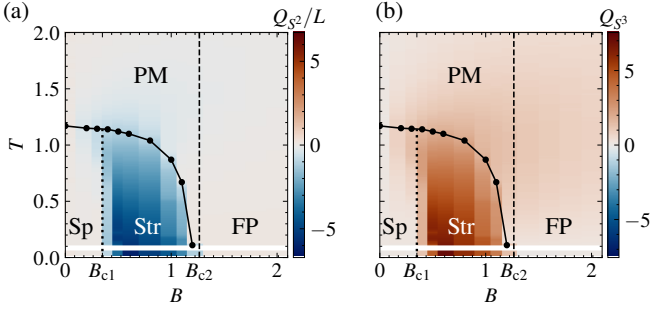
\begin{figure}
\centering
\begin{overpic}[width=\columnwidth]{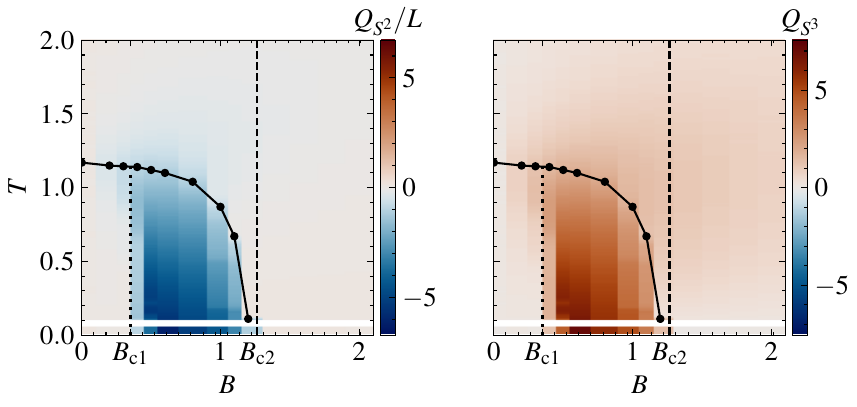}
\put(0,45){(a)}
\put(57,45){(b)}
\put(35,12){\small{FP}}
\put(10,12){\small{Sp}}
\put(20,12){\small{\textcolor{white}{Str}}}
\put(20,35){\small{PM}}
\put(83,12){\small{FP}}
\put(58,12){\small{Sp}}
\put(68,12){\small{\textcolor{white}{Str}}}
\put(68,35){\small{PM}}
\end{overpic}
    \caption{Finite temperature phase diagram for the axially symmetric model with the $\beta$-term in Eq.~\eqref{eq:axial-lattice-model} obtained with fixed twisted boundary conditions for $L=26$, together with $T=0$ iterative minimization results. 
    (a) The absolute value of the $Q_{S^2}/L$ estimates the number of strings in the string lattice phase (Str). Each string carries a half Skyrmion at its extrema, for a total of one Skyrmion per string. The Skyrmion number is given by the topological charge $Q_{S^3}$ in panel (b). 
    The black dots mark specific heat maxima (not in the figure), while the lines represent a guide for the eye. 
    The continuous line tracks the transition temperature, while the dotted line at $0.3\leq B_{\mathrm{c1}}\leq0.4$ follows the drop of the string number for small $B$. At this point, the spin spiral (Sp) becomes energetically favored, in agreement with the prediction of the continuum theory  $B_{\mathrm{c1}}\approx0.37$, see Fig.~\ref{fig:axial-spiral-vs-string}.  
    The dashed vertical line at $B_{\mathrm{c2}}\approx 1.266$ is the expected critical Zeeman field in the continuum theory,
    above which the system reaches the polarized phase. Our numerical data present a transition from the string lattice phase to the field-polarized (FP) phase around the predicted value.
    At high temperatures, the system enters the paramagnetic (PM) phase.
    }
    \label{fig:axial-tbc-pd}
\end{figure}
The finite temperature phase diagram is shown in Fig.~\ref{fig:axial-tbc-pd}. The data for $T>0.1$ are obtained from Monte Carlo simulations, while the data for $T<0.05$ represent the $T=0$ results obtained with iterative minimization. The finite-temperature thickness is used to help visualize the results~\footnote{The black dots identifying the transition temperatures are obtained from specific heat maxima, which are not shown here.}.
At low temperature, the phase diagram shows different regimes depending on the external field strength $B$.
For $B\geq B_{\mathrm{c2}}\approx1.3$, there is a field-polarized (FP) phase, connected with the paramagnetic (PM) phase, with magnetic moments aligned with the external field $\bm{B}$. This regime is associated with vanishing $\pi_2(S^2)$ and $\pi_3(S^3)$ topological charges, see Fig.~\ref{fig:axial-tbc-pd}(a)-(b).

A string lattice phase appears for $T\lesssim 1.25$ and  $B_{\mathrm{c1}}\lesssim B \lesssim B_{c2}$, with $B_{\mathrm{c1}}\approx 0.35$. The string lattice is characterized by the proliferation of strings stretched along the $z$-direction. Each string terminates with half a Skyrmion at its extrema, which is forced to stay inside the lattice by the fixed twisted boundary conditions.
The number of strings at low temperatures is determined by an interplay of string radius size, which depends on the choice of the couplings $\kappa$ and $B$, and the linear size $L$ of the system. Naively, the number of strings is measured by the ratio $|Q_{S^2}|/L$ of the $\pi_2(S^2)$ topological charge defined on the $xy$ plane, see Eq.~\eqref{eq:S2-topological-charge}, and the number of planes along the string direction, which coincides with $L$ in this case. This quantity $Q_{S^2}/L$, increases in absolute value below the transition temperature for $B_{\mathrm{c1}}\lesssim B\lesssim B_{\mathrm{c2}}$, highlighting the onset of a string lattice, see in Fig.~\ref{fig:axial-tbc-pd}(a).
At the same temperature, the topological charge $Q_{S^3}$ defined in Eq.~\eqref{eq:S3-topological-charge}, which measures the number of Skyrmions, increases in magnitude, as reported in Fig.~\ref{fig:axial-tbc-pd}(b). Approximatively, we see that $|Q_{S^3}|\approx |Q_{S^2}|/L$, corroborating our picture of each string carrying half a Skyrmion at its extrema, for a total of one Skyrmion per string.

\begin{figure}
    \centering
    \begin{overpic}[width=\columnwidth]{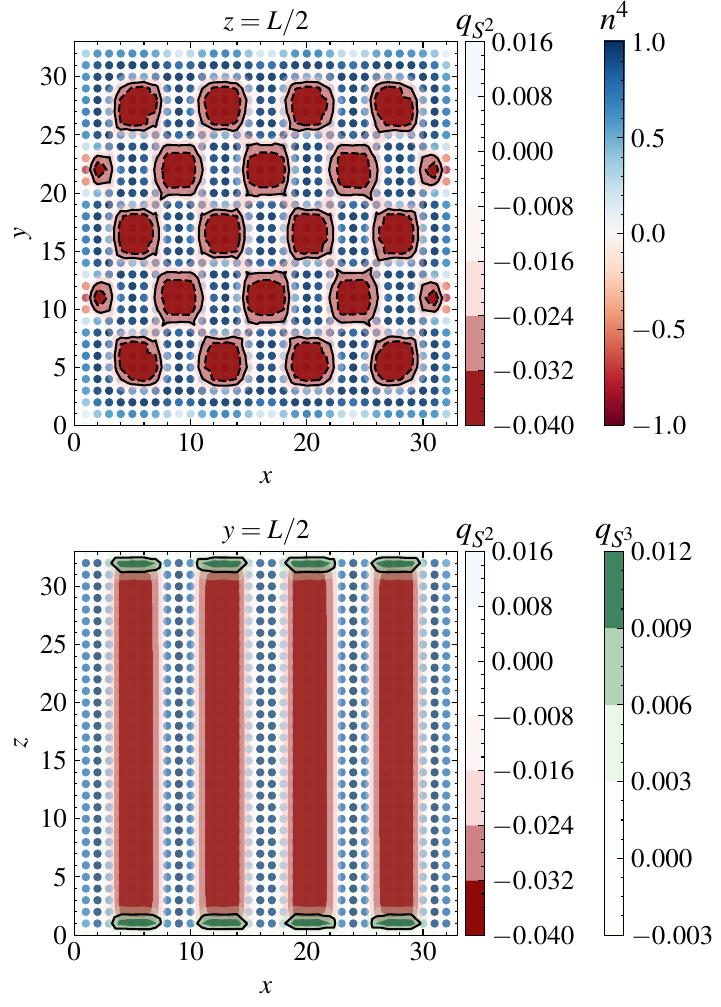}
    \put(8,97){(a)}
    \put(8,46){(b)}
    \end{overpic}
    \caption{Distribution of the $\pi_2(S^2)$ topological charge and $\pi_3(S^3)$ topological charge per site in a zero temperature configuration for $L=34$ and $B=0.6$.
    (a) The cut at $z=L/2$ exposes a topological charge per site $q_{S^2}$ arranged in a triangular pattern, corresponding to the string lattice.
    (b) The stretch of the strings along the $z$-direction is suggested by the cut at $y=L/2$. In this case, $q_{S^2}$ fills cylinders along the $z$-direction, corresponding to the strings. Each string has a concentration of $\pi_3(S^3)$ topological charge at its ends, representing the splitting of a Skyrmion into two halves and detached by the string in between them. The background spins not involved in the string lattice align along the vacuum direction (0,0,0,1).}
    \label{fig:axial-tbc-zero-T-charge-densities}
\end{figure}

The strings and the associated Skyrmions are exposed with the ground state configuration obtained with iterative minimization. 
We measured the $\pi_2(S^2)$ and $\pi_3(S^3)$ topological charges per site as in Eq.~\eqref{eq:charge-and-widning-densities}. The results are displayed in Fig.~\ref{fig:axial-tbc-zero-T-charge-densities} for the representative case $L=34$ and $B=0.6$.
Specifically, the cut at $z=L/2$ in Fig.~\ref{fig:axial-tbc-zero-T-charge-densities}(a), shows a triangular lattice pattern of almost circular disks where the $q_{S^2}$ topological charge density is concentrated. This emerging lattice develops in a background of spins aligned along the vacuum direction $(0,0,0,1)$.
The same configuration is shown in Fig.~\ref{fig:axial-tbc-zero-T-charge-densities}(b), for a cut at $y=L/2$. This panel identifies the $z$-axial direction of the strings, with parallel cylinders of $q_{S^2}$ finite density. The ends of the strings feature two regions of $q_{S^3}$ associated with the $\pi_3(S^3)$ topological charge per site. These  ``pancake''-like regions occupy the same transverse area of the strings, and are highlighted in green in Fig.~\ref{fig:axial-tbc-zero-T-charge-densities}(b). The accumulation of the $\pi_3(S^3)$ topological charge at the string extrema is predicted by the continuum theory in Sec.~\ref{sec:beta_continuum_theory}, and represents the splitting of a Skyrmion into two halves, separated by the string.
The string number approaches zero as the Zeeman field increases towards $B\approx1.3$. From a continuum point of view, the string energy becomes positive above a certain critical field $B_{\mathrm{c2}}$, and the strings and associated Skyrmions disappear. With the choice of coupling $\kappa=\sqrt{2}$, the continuum theory predicts $B_{\mathrm{c2}}\approx 1.6$ above which the field-polarized phase becomes favorable.
The lattice calculations report a critical field that has a slightly lower value than the continuum estimate of $B_{\mathrm{c2}}$, as the string phase disappears around $B_{\mathrm{c2}}\approx1.3$, see Fig.~\ref{fig:axial-tbc-pd}.

\begin{figure}
    \centering
    \includegraphics[width=0.9\linewidth]{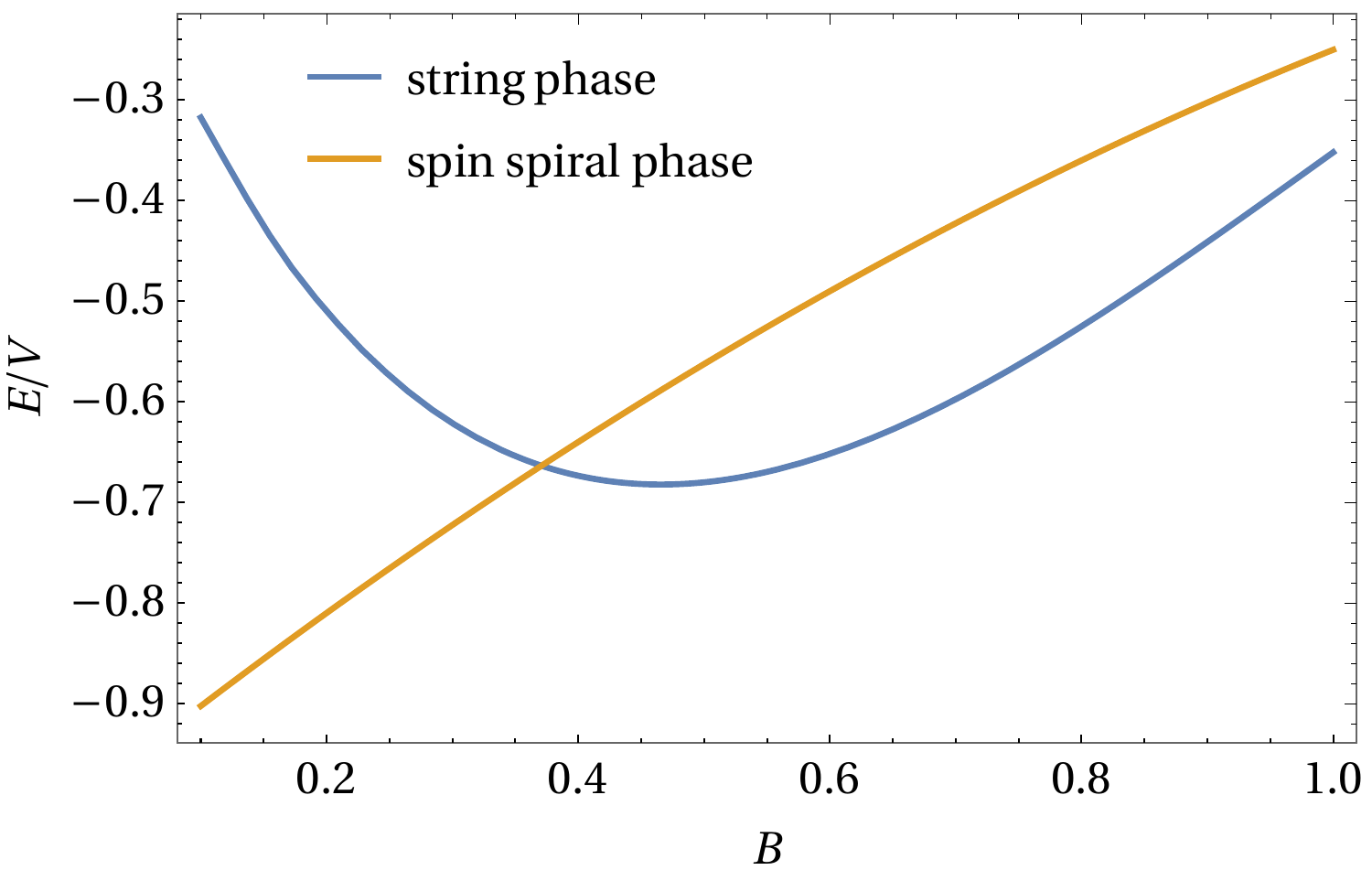}
    \caption{Energy density comparison between a spin-spiral phase and a string lattice phase as a function of the Zeeman field $B$. The spin-spiral state is favored at low fields, up to $B\approx0.37$, at which the string lattice becomes the lowest energy state. 
    In this figure, the energy density of the string phase is estimated from a single string's energy and size. }
    \label{fig:axial-spiral-vs-string}
\end{figure}

In the low-field regime for $B\lesssim B_{\mathrm{c1}}$, the string number suddenly decreases. This could be due to two interplaying factors. 
As the Zeeman term decreases, the string radius enlarges. On a finite-size system, it becomes impossible to accommodate the strings since they are too large. 
However, there is also an intrinsic energetic competition in the bulk between a spin-spiral state, realized at $\bm{B}=0$, see App. \ref{subapp:axialDMI-LT}, and the string lattice. For small Zeeman fields, the spin spiral is energetically favored, and the string lattice is destroyed. 
This scenario is confirmed by calculations in the continuum: it is possible to approximately calculate the energy of the spin-spiral state in the presence of an external Zeeman field and compare it to the string lattice energy. The approximation considers an average spin-spiral state on its periodic cell, so that several oscillating terms are neglected in the minimization problem. 
As a consequence, the strong constraint on the magnetic moment $\bm{n}(\bm{r})\cdot\bm{n}(\bm{r})=1$ is replaced with the spatially averaged version $\langle \bm{n}(\bm{r})\cdot \bm{n}(\bm{r}) \rangle = 1$. Within the approximation, the spin-spiral energy density is given by the expression
\begin{equation}
    \label{eq:spin-spiral-approx-energy}
    \mathcal{E}=-\frac{Ja^{-3}}{2}\left(\kappa-\frac{B}{D} \right)^2\,,
\end{equation}
which is used to calculate the energy in a box of given volume. Then, the spin-spiral energy is compared with the string lattice energy. The lattice is assumed to host a number of strings given by the total area of a face of the box divided by the area of a single string, without any distortions. 
The comparison between the two energy densities is shown in Fig.~\ref{fig:axial-spiral-vs-string}. Within these approximations, we find $B_{\mathrm{c1}}\approx 0.37$, which is close to the value obtained on the lattice for which the string number drops to zero, see the dotted line in Fig.~\ref{fig:axial-tbc-pd}(a). Details on the approximation of the spin-spiral energy density are given in App.~\ref{appendix:spin-spiral-continuum-method}.
%
%
\subsection{Lattice theory with the \texorpdfstring{$\alpha$}{alpha}-term}   %
\label{subsec:spherical-DM-numerics} 
The model with the $\alpha$-term in Sec.~\ref{sec:genDM}, has a spherically invariant DMI term, see Eq.~\eqref{eq:DM_Neel}. This term is discretized on the lattice as in Eq.~\eqref{Neeldiscrete}, and together with the Heisenberg contribution in Eq.~\eqref{Heisenberg_discrete} and the Zeeman interaction in Eq.~\eqref{Zeeman_discrete}, leads to an Hamiltonian
\begin{equation}
    \label{eq:spherical-lattice-model}
    \begin{split}
        \mathcal{H} = &-J\sum_{I,\mu}\bm{n}^{(I)}\cdot \bm{n}^{(I+a\hat{\bm{e}}_\mu)} \\
        &+D\sum_{I,\mu}\bigl[
            (\hat{\bm{e}}_4 \cdot \bm{n}^{(I)})
            (\hat{\bm{e}}_\mu\cdot \bm{n}^{(I+a\hat{\bm{e}}_\mu)})\\
        &\hphantom{{}+D\sum_{I,\mu}\bigl[}
            -(\hat{\bm{e}}_\mu\cdot\bm{n}^{(I)})
            (\hat{\bm{e}}_4\cdot\bm{n}^{(I+a\hat{\bm{e}}_\mu)})
        \bigr] \\
        &-\bm{B}\cdot \sum_{I}\bm{n}^{(I)}\,,
    \end{split}
\end{equation}
where the unit vectors $(\hat{\bm{e}}_\mu)_\alpha = \delta_{\mu\alpha}$ for $\mu=x,y,z$ generalize the three-dimensional cubic vectors with an additional fourth component fixed to zero. In this case, the unit vector $\hat{\bm{e}}_4=(0,0,0,1)$ specifies the vacuum direction, since the Zeeman field is $\bm{B}=B\hat{\bm{e}}_4$. 
The couplings are fixed to $J=1$ and $D=\sqrt{3}$, meaning $\kappa=\sqrt{3}$. In the $\bm{B}=0$ case and with periodic boundary conditions, this choice of values yields a spin-spiral ground state with known wavevector $\bm{K}$, as described in App.~\ref{subapp:spherical-DMI-LT}.
Different from the axial case, here we considered periodic boundary conditions.
\begin{figure}
    \centering
    \begin{overpic}[width=\columnwidth]{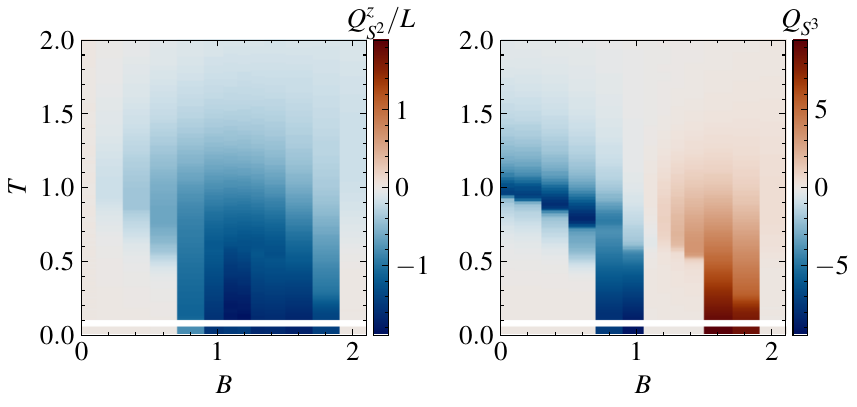}
        \put(0,45){(a)}
        \put(59,45){(b)}
        \put(60,12){\small{Sp}}
        \put(69.5,12){\textcolor{white}{\small{aSk}}}
        \put(76.5,12){\small{Str}}
        \put(83.5,12){\textcolor{white}{\small{Sk}}}
        \put(80,38){\small{PM}}
    \end{overpic}
    \caption{Finite-temperature phase diagram for the (a) $\pi_2(S^2)$ topological charge and (b) $\pi_3(S^3)$ topological charge for the case with the $\alpha$-term, together with zero-temperature iterative minimization results for $L=16$ and periodic boundary conditions.
    For large fields, $B>B_{\mathrm{c4}}$, the system is fully polarized at low temperatures. This phase is topologically trivial, with vanishing $\pi_2(S^2)$ and $\pi_3(S^3)$ topological charges.
    As the field decreases from large values, the system enters a low-temperature Skyrmion lattice phase, highlighted by the positive $\pi_3(S^3)$ topological charge, see red region in panel (b) for $B_{\mathrm{c3}} \leq B\leq B_{\mathrm{c4}}$. The skymions also carry $\pi_2(S^2)$ topological charge, which appears in blue in panel (a).
    The system shifts to a different phase if the field further decreases, $B_{\mathrm{c2}} \leq B\leq B_{\mathrm{c3}}$. In particular, the Skyrmion lattice phase vanishes, as marked by the vanishing $\pi_3(S^3)$ topological charge. However, there is still a finite $\pi_2(S^2)$ topological charge associated with a string lattice phase.
    At slightly lower fields, $B_{\mathrm{c1}} \leq B\leq B_{\mathrm{c2}}$, the system is driven into a different phase before rearranging into the spin-spiral phase. This phase is characterized by negative and finite $\pi_2(S^2)$ and  $\pi_3(S^3)$ topological charges, blue regions in panels (a) and (b). In this regime, an antiSkymrion lattice phase is realized.  %
    Below the lower critical field, $B<B_{\mathrm{c1}}$, the low-temperature stable phase becomes a spin spiral with vanishing $\pi_3(S^3)$ and $\pi_2(S^2)$ topological charges.
    }
    \label{fig:spherical-PD}
\end{figure}
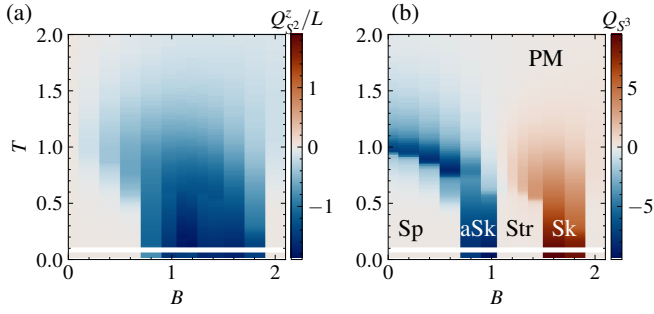

The finite temperature phase diagram together with the zero-temperature minimization results are shown in Fig.~\ref{fig:spherical-PD}. Many phases appear as the external field is varied. 
At a large field, $B\geq B_{\mathrm{c4}}\approx 1.9$, the system is fully polarized at low temperature, with magnetic moments pointing towards the $\hat{\bm{e}}_4$-direction. From the continuum theory, the Skyrmion solution is stabilized for $\tilde{\kappa}>\sqrt{3}$, see Eq.~\eqref{eq:Rpm_magskyrm_sphaleron}. The parameter $\tilde{\kappa}=\kappa\sqrt{2/JB}\propto B^{-1/2}$, meaning that the Skyrmion ceases to exist above a critical field $B_{\mathrm{c4}}=2$ for our choice of couplings $\kappa=\sqrt{3}$ and $J=1$, which agrees with the Monte Carlo data in Fig.~\ref{fig:spherical-PD}.
Conversely, for $B<B_{\mathrm{c1}}\approx0.7$, the system is stabilized into a spin-spiral state, with a single wavevector $\bm{K}$. Details on the spin-spiral state are given in App.~\ref{subapp:spherical-DMI-LT}.
The two limits of high and low fields correspond to phases of trivial topology: at low temperatures, the $\pi_2(S^2)$ topological charge and the topological charge associated with the homotopy group $\pi_3(S^3)$ are exactly zero, as shown in Fig~\ref{fig:spherical-PD}.
\subsubsection{3D spherical Skyrmion lattice}
\begin{figure}
    \centering
    \begin{overpic}[width=\columnwidth]{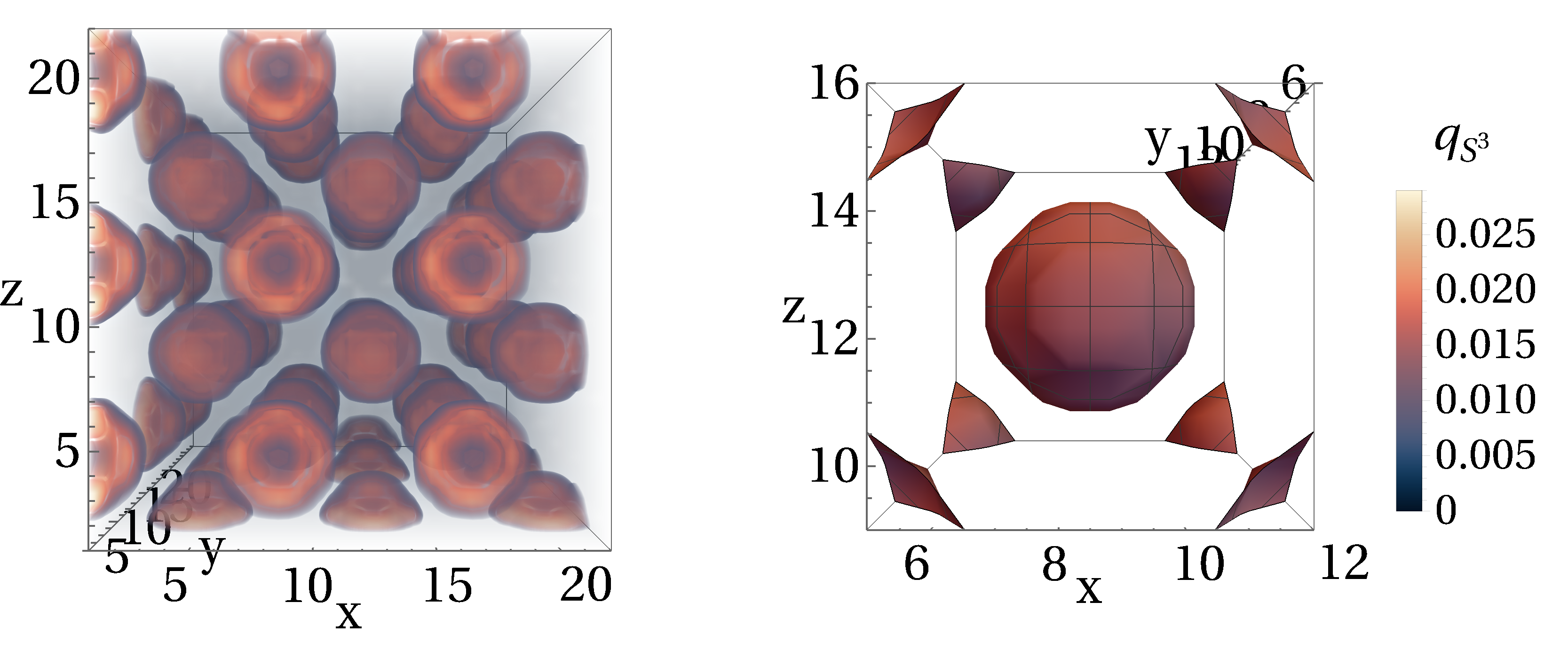}
        \put(0,45){(a)}
        \put(48,45){(b)}
    \end{overpic}
    \caption{(a) Emergent Skyrmion lattice obtained with iterative minimization for the representative case $B=1.6$ and $L=24$. (b) Zoom of the lattice considering an isosurface at $q_{S^3}=0.012$. The Skyrmions form a body-centered cubic lattice with a two-Skyrmion unit cell.}
    \label{fig:spherical-Skyrmion-lattice}
\end{figure}

As the field decreases from large values, the system enters a Skyrmion lattice phase, highlighted by the positive-charge region around $B_{\mathrm{c3}} \lesssim B \lesssim B_{\mathrm{c4}}$ in Fig.~\ref{fig:spherical-PD}(b), with $B_{\mathrm{c3}}\approx 1.5$. 
This phase consists of an emergent lattice formed by spherical Skyrmions, as shown in Fig.~\ref{fig:spherical-Skyrmion-lattice}(a). The Skyrmions arrange themselves in a body-centered cubic lattice with a two-Skyrmion unit cell, see  Fig.~\ref{fig:spherical-Skyrmion-lattice}(b), and have a radius between two and three sites.

\begin{figure}
    \centering
    \begin{overpic}[width=\columnwidth]{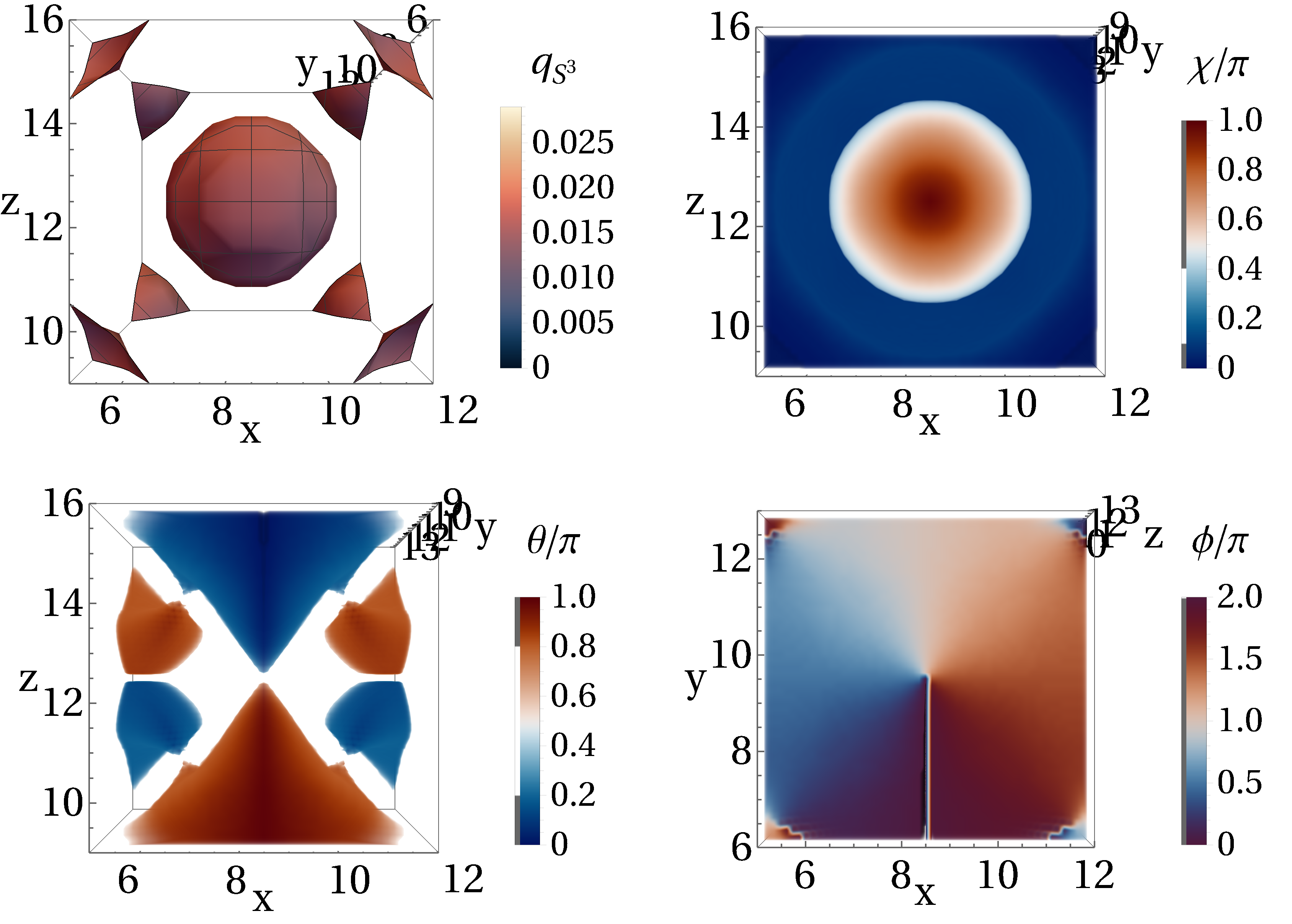}
        \put(-5,70){(a)}
        \put(50,70){(b)}
        \put(-5,30){(c)}
        \put(50,30){(d)}
    \end{overpic}
    \caption{(a) Single Skyrmion of the BCC Skyrmion lattice,  and cross sections (b)-(c)-(d) through its center to reveal the phases of the magnetic moments. (b) Cross section at constant $y$ through the Skyrmion center. The angle $\chi$ defines the profile of the Skyrmion: in the center, the magnetic moment is antiparallel to the vacuum, as evidenced by $\chi=\pi$, while outside the Skyrmion the moments align along the vacuum, with $\chi=0$.
    (c) The same cross section is used to expose the phase $\theta$, which changes from $\theta=0$ to $\theta=\pi$, going from the north pole to the south pole of the Skyrmion, in agreement with the hedgehog Ansatz of Eq.~\eqref{eq:hedgehog}. 
    (d) Phase $\phi$ for a cross section at constant $z$ through the Skyrmion equator. The angle $\phi$ winds precisely one time around the Skyrmion.}
    \label{fig:spherical-Skyrmion-angles}
\end{figure}
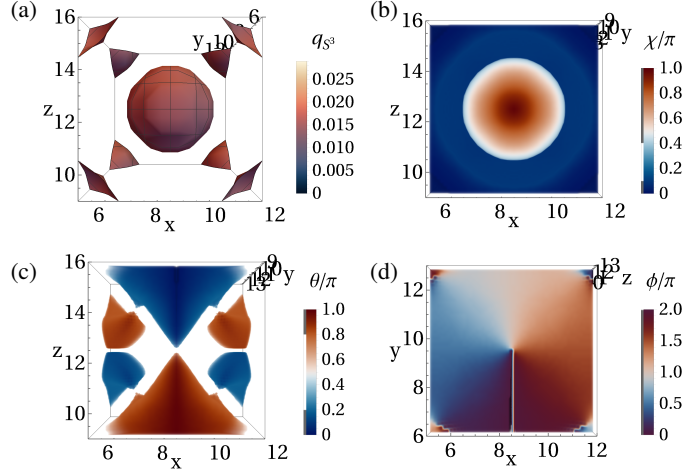

Three angles characterize each magnetic moment. In the Skyrmion lattice phase, these angles are distributed in a specific way. Figure~\ref{fig:spherical-Skyrmion-angles} shows the phases $(\phi,\theta,\chi)$, defined in Eq.~\eqref{eq:hedgehog}, for a single Skyrmion obtained from a minimized configuration with $B=1.6$ and $L=24$. 
The phase $\chi$ encodes the Skyrmion profile, with $\chi=\pi$ at the center of the Skyrmion corresponding to a moment with opposite direction with respect to the vacuum, while $\chi=0$ on its outside marks the moments parallel to the vacuum, see Fig.~\ref{fig:spherical-Skyrmion-angles}(b).
The angle $\theta$ varies from $\theta=0$ to $\theta=\pi$, going from the north pole to the south pole of the Skyrmion, as shown in Fig.~\ref{fig:spherical-Skyrmion-angles}(c).
On the other hand, the phase $\phi$ winds precisely one time around the spherical Skyrmion, see Fig.~\ref{fig:spherical-Skyrmion-angles}(d).
The phases behave as expected from the hedgehog Ansatz in the continuum theory, see Sec.~\ref{sec:genDM}. Hence, the discrete lattice problem exhibits an emerging BCC Skyrmion lattice of 3D spherical Skyrmions.

\begin{figure}
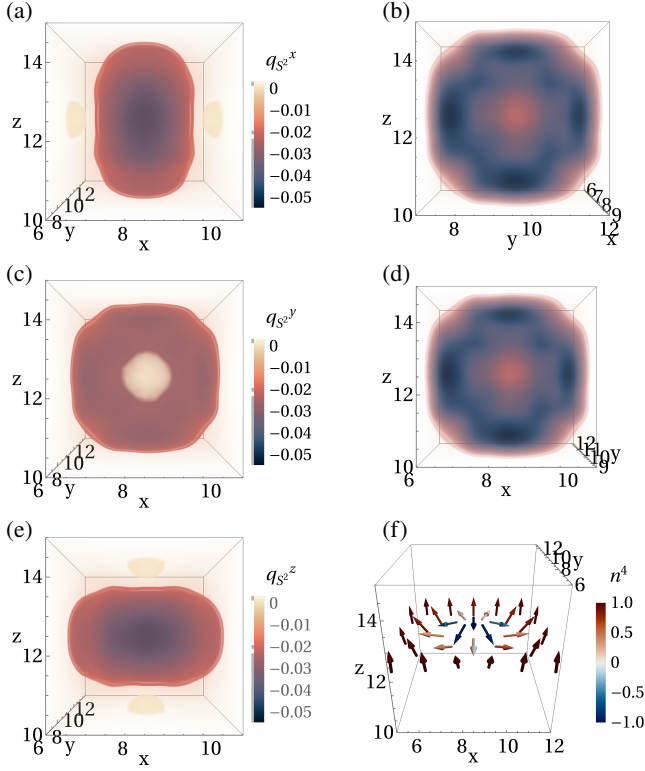

    \centering
    \begin{overpic}[width=\columnwidth]{spherical_pbc_B1.6_L24_q2charges.png}
        \put(0,100){(a)}
        \put(0,65){(c)}
        \put(0,31){(e)}
        \put(50,100){(b)}
        \put(50,65){(d)}
        \put(50,31){(f)}
    \end{overpic}
    \caption{The three $\pi_2(S^2)$ charge densities in correspondence to the Skyrmion in Fig.~\ref{fig:spherical-Skyrmion-angles}(a), forming the charge vector $\bm{q}_{S^2}$. 
    The $x$ component is shown in panel (a): $q_{S^2}^x$ is distributed on a pancake-like shape within the Skyrmion. The cross section of the Skyrmion's center at constant $x$ in panel (b) reveals the areas of maximal charge. An integration on this plane returns a $\pi_2(S^2)$ charge $Q_{S^2}^x\approx-0.93$. 
    Same plots (c)-(d) for the component $q_{S^2}^y$. However, the charge distribution is rotated by $\pi/2$ degrees in the $xy$ plane. Numerical integration of the charge density gives $Q_{S^2}^y\approx-0.93$. 
    The distribution of $q_{S^2}^z$ is displayed in panel (e). A cross section of the Skyrmion's center at constant $z$ reveals a 2D Néel Skyrmion, see panel (f), which comes from the hedgehog Ansatz of Sec.~\ref{sec:genDM} evaluated at $\theta=\pi/2$. Numerical integration of the charge yields $Q_{S^2}^z\approx-0.93$ also in this case.}
    \label{fig:spherical-q2-charges}
\end{figure}

The topological properties are summarized in the $\pi_3(S^3)$ topological charge carried by a single Skyrmion. We measured~\footnote{We used an interpolation procedure on the lattice magnetic moments to reduce the discretization effects.} the $Q_{S^3}$ charge of a single Skyrmion obtained in our simulations, finding $Q_{S^3}\approx0.97$, very close to the expected unit charge for the hedgehog Skyrmion described by the continuum theory in Eq.~\eqref{eq:hedgehog}.
The spherical Skyrmion also carries a topological charge related to the homotopy group $\pi_2(S^2)$. In particular, it is possible to define a three-component vector of $\pi_2(S^2)$ charge densities, or fluxes~\cite{azhar24}, $\bm{q}_{S^2}$ which encodes non-trivial winding on the cartesian planes of the effective magnetic moments $\bm{n}^{(x)}=(n_2,n_3,n_4)$, $\bm{n}^{(y)}=(-n_1,n_3,n_4)$, and $\bm{n}^{(z)}=(n_1,n_2,n_4)$.
Figure~\ref{fig:spherical-q2-charges} shows the density plot of the components of $\bm{q}_{S^2}$, together with a cut at constant $x$, $y$, or $z$ passing through the Skyrmion center. The $\pi_2(S^2)$ charges concentrate within the Skyrmion, as visible from the three-dimensional plots in Fig.~\ref{fig:spherical-q2-charges}(a)-(c)-(e), and the respective cross sections at constant $x,y,z$ through the center of the Skyrmion, in Fig.~\ref{fig:spherical-q2-charges}(b)-(d)-(f). In particular, Figure~\ref{fig:spherical-q2-charges}(f) shows the spin configuration on a plane at constant $z$ for the effective moments $\bm{n}^{(z)}=(n_1,n_2,n_4)$, with the spins arranged in a typical $\pi_2(S^2)$ hedgehog configuration, known also as 2D N\'eel Skyrmion. This configuration comes from the hedgehog Ansatz in Sec.~\ref{sec:genDM} evaluated at $\theta=\pi/2$, and on the lattice corresponds to the configuration of the moments $\bm{n}^{(z)}$ on the plane at constant $z$ passing through the Skyrmion equator.
On a single Skyrmion, the integral 
\begin{equation}
    \label{eq:q2-charges}
    Q_{S^2}^j = \int_\Sigma \mathrm{d}\boldsymbol{\Sigma}^j\cdot \bm{q}_{S^2}
\end{equation}
gives the  $\pi_2(S^2)$ topological charge associated with the Skyrmion, and on the plane at constant $j$~\footnote{For example, $Q_{S^2}^z$ is obtained with integration on the $xy$ plane at constant $z$, with $\mathrm{d}\boldsymbol{\Sigma}^z=(0,0,\mathrm{d}x\mathrm{d}y)$.}. In our case, we found $Q^x_{S^2}=Q^y_{S^2}=Q^z_{S^2}=-0.93$, which is close to the expected unit magnitude.
The cubic symmetry acts on both orbital and spin spaces, and as a consequence, it permutes the components of the $\bm{q}_{S^2}$ vector. Hence, although the single components of $\bm{q}_{S^2}$ have lower symmetry, the configuration of the vector $\bm{q}_{S^2}$ is invariant under cubic symmetry, since the three components of $\bm{q}_{S^2}$ are the same.
Overall, the lattice Skyrmion matches the hedgehog Skyrmion of the continuum theory described in Sec.~\ref{sec:genDM}.

\subsubsection{String lattice}
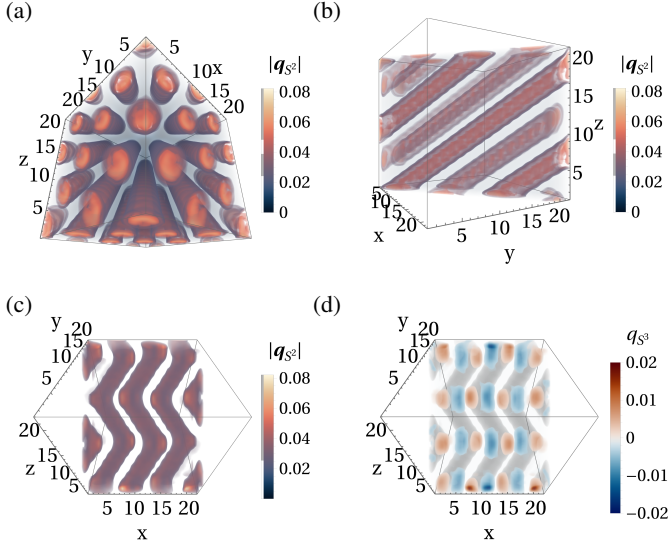
\begin{figure}
    \centering
    \begin{overpic}[width=\columnwidth]{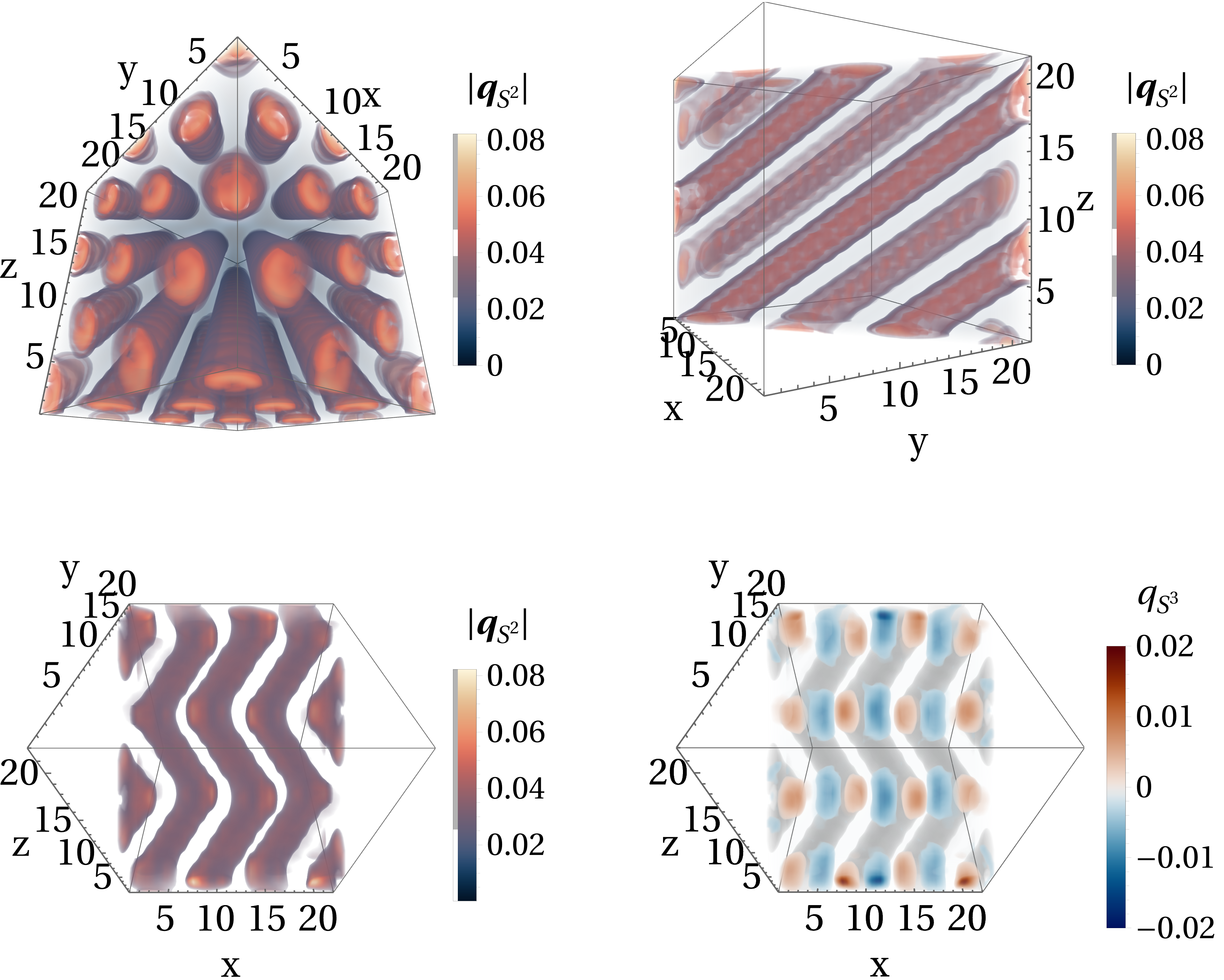}
        \put(-2,80){(a)}
        \put(45,80){(b)}
        \put(-2,35){(c)}
        \put(45,35){(d)}
    \end{overpic}
    \caption{Two types of string lattice obtained for $L=24$ and (a)-(b) $B=1.2$, (c)-(d) $B=1.1$.
    (a) The straight strings are directed along [111] direction.
    (b) Strings are exposed by cuts with $-2.5\leq x+y\leq 4$.
    (c) Representative example for zigzag strings along the [011] direction. Only one plane is shown; however, the string lattice features stacked planes with the same features as the ones shown here. The strings
    oscillate around the average propagation direction.
    (d) The regions of direction-change are associated with a fractional $\pi_3(S^3)$ topological charge. Since the angle of the string knees is around $2\pi/3$, the expected charge is $Q_{S^3}=\pm 1/6$, which agrees with the numerical integration $Q_{S^3}\approx\pm0.19$.
    Since it splits into positive and negative values, the global $\pi_3(S^3)$ charge is zero.}
    \label{fig:string-representative-cases}
\end{figure}

If the field decreases below a certain threshold $B_{\mathrm{c3}}$, the Skyrmion lattice disappears. The phase diagram in Fig.~\ref{fig:spherical-PD}(a)-(b) shows a region of vanishing $\pi_3(S^3)$ topological charge in the range $1.1\lesssim B\lesssim1.5$, with a simultaneous finite $\pi_2(S^2)$ topological charge.
Inspection of the low-temperature spin configurations reveals a string lattice phase. From the numerical calculations, we found two types of string ordering: straight or zigzag, depending on the field values and lattice size. 
Two representative cases are shown in Fig.~\ref{fig:string-representative-cases}. The straight strings of Fig.~\ref{fig:string-representative-cases}(a)-(b) are directed along the [111] direction and are obtained for $1.2\lesssim B \lesssim B_{\mathrm{c_3}}$, while the zigzag strings of Fig.~\ref{fig:string-representative-cases}(c)-(d) seem to appear for lower values of the Zeeman field, $B_{\mathrm{c2}}\lesssim B\lesssim1.2$. 
In this scenario, the strings develop a wave-like modulation around their average propagating direction, as shown in Fig.~\ref{fig:string-representative-cases}(c)-(d) for representative strings along the [011] direction for $B=1.1$. 
There is a $\pi_3(S^3)$ topological charge distribution in correspondence with the point where the string changes direction. However, it splits into pairs of positive and negative values, so the global $S^3$ charge remains zero, see Fig.~\ref{fig:string-representative-cases}(d).
This phenomenon can be explained with a topological argument.
Qualitatively, we can understand the presence of a fractional $S^3$ charge by a ``knee'' on an $S^2$-charged string.
The $S^2$-charged string is a map from the plane to the two-sphere and can be viewed as a global vortex or an $S^2$ Skyrmion.
Considering the string on a half-plane, we can close the string in on itself, by rotating around the axis of the half-plane, forming a ring.
Topology dictates that the $S^3$-charge is an integer and is given by the number of ``twists'' of the string: the simplest nontrivial case is a single twist giving $Q_{S^3}=\pm1$.
Topologically, we can deform the ring to a triangle, a square or an $n$-gon with each corner now containing a localized $S^3$ charge of $1/n$.
Since the ``knee'' seen in Fig.~\ref{fig:string-representative-cases}(c)-(d) bends by approximately 120$^\circ$, we expect the charge to be $Q_{S^3}=\pm1/6$ if it is nonvanishing~\footnote{On the lattice, the angle is around $116^\circ$.}.  
We measured the charges on the lattice, obtaining $|Q_{S^3}|\approx0.19$, which is quite close to the expected value.
%

\subsubsection{AntiSkyrmion lattice}
Before reaching the spin-spiral phase at low fields, the system enters a different regime, characterized by negative $\pi_3(S^3)$ and $\pi_2(S^2)$ charges, see Fig.~\ref{fig:spherical-PD}(b) in the range $B_{\mathrm{c1}}\lesssim B\lesssim B_{\mathrm{c2}}$. 
\begin{figure}
    \centering
    \begin{overpic}[width=\columnwidth]{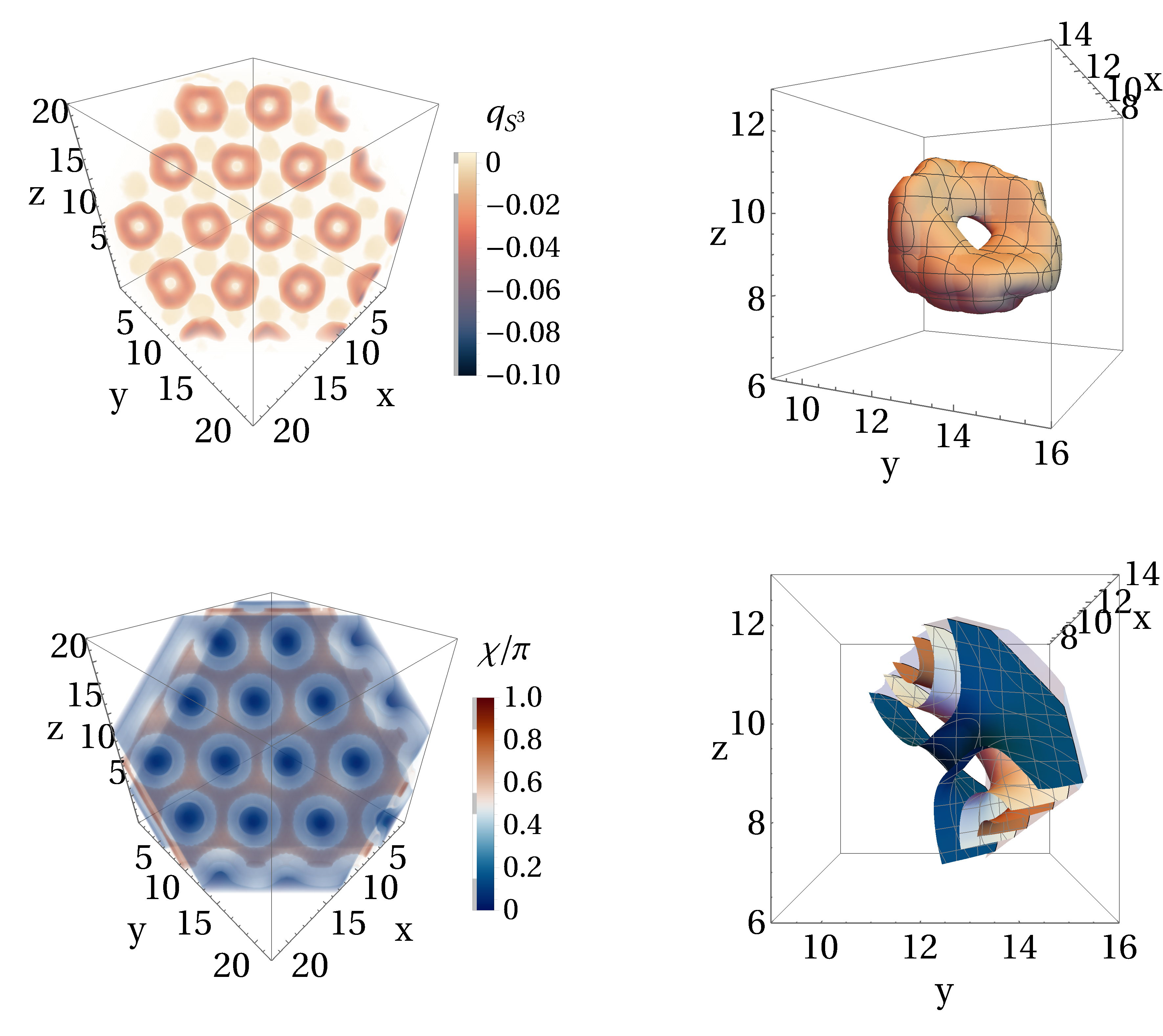}
        \put(0,85){(a)}
        \put(60,85){(b)}
        \put(0,40){(c)}
        \put(60,40){(d)}
    \end{overpic}
    \caption{AntiSkyrmion lattice for $B=0.8$ and $L=24$. The antiSkyrmions are distributed on planes with triangular patterns, with parallel planes along the [111] direction.
    In panel (a), we show a representative plane, with the topological charge density $q_{S^3}$ forming tori rings. 
    (b) A zoom of the isosurface at $q_{S^3}=-0.015$ exposes the torus shape of the antiSkyrmion.
    (c) The analysis of the $\chi$ angle reveals the structure of the $n^4$ component in the antiSkyrmion, with moments parallel to the vacuum passing through the antiSkyrmion center, and moments antiparallel to the vacuum forming a net on the antiSkyrmion plane.
    (d) A zoom on the antiSkyrmion shows the structure of $\chi$ for three isosurfaces at $\chi=0.15$ (blue), $\chi=0.5$ (white), $\chi=0.75$ (red).
    }
    \label{fig:spherical-antiSkyrmion-lattice}
\end{figure}
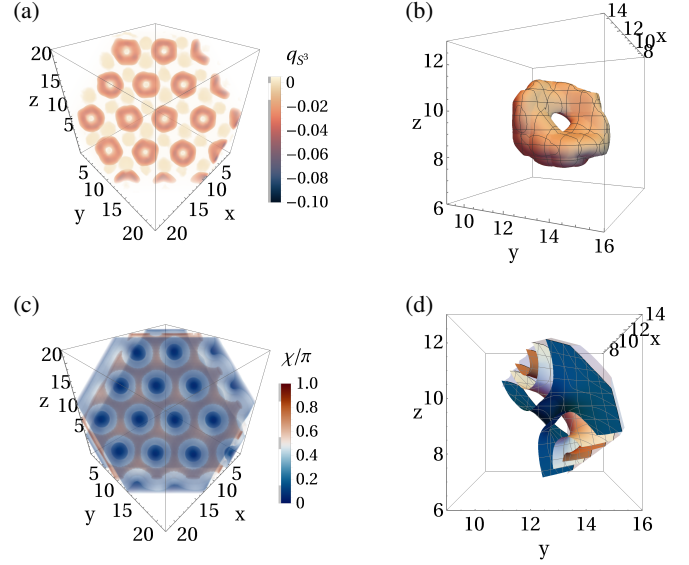
This phase presents a triangular pattern in the distribution of the $\pi_3(S^3)$ charge on parallel planes along the [111] direction. In Figure~\ref{fig:spherical-antiSkyrmion-lattice}(a), we show one isolated plane on the lattice. We can identify single units with torus shapes for the $S^3$ charges, as shown in Fig.~\ref{fig:spherical-antiSkyrmion-lattice}(b). A single torus carries a $\pi_3(S^3)$ topological charge $Q_{S^3}\approx-0.99$, so it is reasonable to associate those objects with unit-charged antiSkyrmions.
The phase $\chi$ exposes an atypical pattern on the lattice. Different from the Skyrmion or string lattice case, where the region of $\chi=\pi$ is bounded within the Skyrmion ball or within the strings, we have a pattern of $\chi=\pi$ which extends on entire planes on the lattice, as marked by the red region in Fig.~\ref{fig:spherical-antiSkyrmion-lattice}(c). 
The region with antiparallel moments is penetrated by holes with moments aligned along the vacuum, forming the triangular pattern shown in Fig.~\ref{fig:spherical-antiSkyrmion-lattice}(c) and a zoom around the antiSkyrmion region in Fig.~\ref{fig:spherical-antiSkyrmion-lattice}(d). It is around these holes that the $\pi_3(S^3)$ charge is concentrated, as highlighted by Fig~\ref{fig:spherical-antiSkyrmion-lattice}(b)-(d).

\begin{figure}
    \centering
    \begin{overpic}[width=\columnwidth]{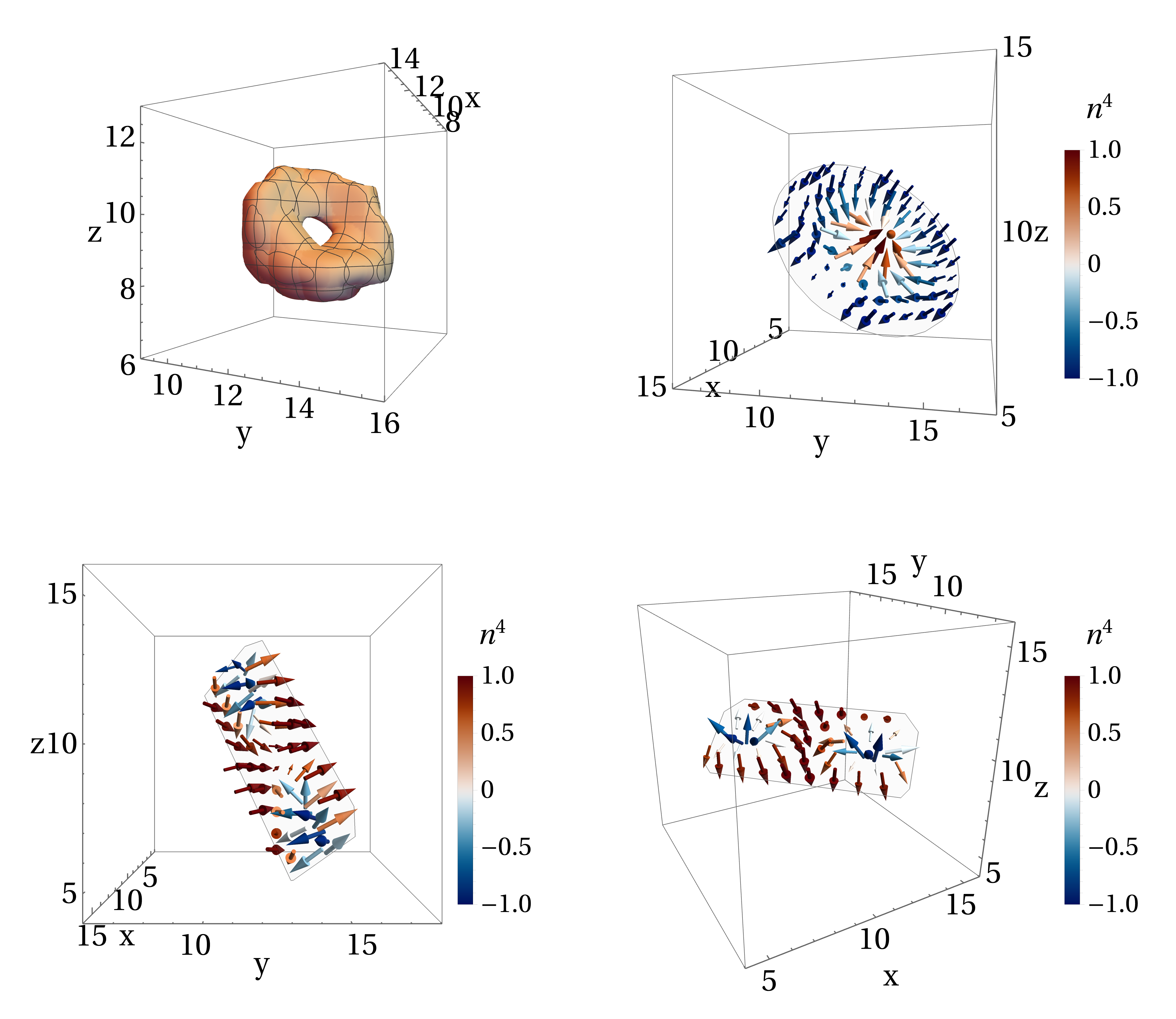}
        \put(0,80){(a)}
        \put(50,80){(b)}
        \put(0,40){(c)}
        \put(50,40){(d)}
    \end{overpic}
    \caption{(a) Zoom on a single torus-shaped antiSkyrmion. 
    (b) A cut with a plane defined by $x+y+z=32$. The plot shows the effective magnetic moments $(n_a,n_b,n_4)$, where the components are defined in Eq.~\eqref{eq:antiskyrmion-spin-components}. The moments are arranged in an $S^2$ Bloch Skyrmion.
    (c) Cutting with the plane $-\frac{x}{\sqrt{2}}+\frac{y}{\sqrt{2}}=\frac{3}{2}$ through the antiSkyrmion center reveals the $S^2$ Skyrmion inside the torus ring with the effective moments $(n_a,n_c,n_4)$.
    (d) A similar $S^2$ Skyrmion situation arises with a cut with $\frac{x}{\sqrt{6}}+\frac{y}{\sqrt{6}}-\frac{2z}{\sqrt{6}}=\frac{3}{2}$ and effective moments $(n_b,n_c,n_4)$.}
    \label{fig:antiskyrmion-components}
\end{figure}
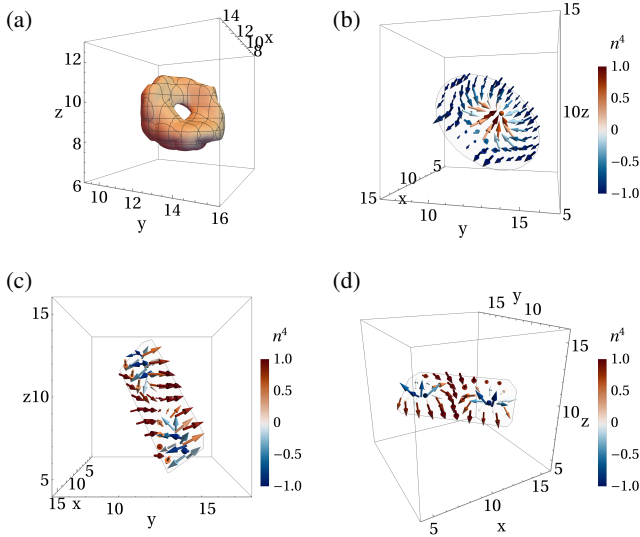

The theory in Eq.~\eqref{eq:spherical-lattice-model} is chiral, implying that the Skyrmion and the antiSkyrmion have different properties. In particular, the antiSkyrmion loses the spherical symmetry and retains an axial symmetry around the [111] axis.
The antiSkyrmion configuration is exposed not only by looking into the $n^4$ component, or equivalently, the $\chi$ phase in Fig.~\ref{fig:spherical-antiSkyrmion-lattice}(c)-(d), but also by revealing the behavior of its effective lower-dimensional magnetic moments.
This behavior is shown in Fig.~\ref{fig:antiskyrmion-components}. By cutting the torus shape of the antiSkyrmion, see Fig.~\ref{fig:antiskyrmion-components}(a), we reveal the configuration of effective magnetic moments. First, we consider a cross section in the [111] plane, splitting the antiSkyrmion into two halves, as shown in Fig.~\ref{fig:antiskyrmion-components}(b), revealing an $S^2$ Bloch Skyrmion in the effective magnetization $\tilde{\bm{n}}=(n_a,n_b,n_4)$~\footnote{The $n_4$ component is always associated to the direction orthogonal to the plane of the cut in Fig.~\ref{fig:antiskyrmion-components}.}. The components are given by
\begin{align}
    \label{eq:antiskyrmion-spin-components}
    n_a&=\bm{n}\cdot\bm{a}\,,\quad n_b=\bm{n}\cdot\bm{b}\,\quad n_c=\bm{n}\cdot \bm c\,, \\
    \bm{a}&=\frac{1}{\sqrt{6}}\begin{pmatrix}
        1 \\ 1 \\ -2 \\ 0
    \end{pmatrix}\,,\quad 
    \bm b= \frac{1}{\sqrt{2}}\begin{pmatrix}
        -1 \\ 1 \\ 0 \\ 0
    \end{pmatrix}\,, \quad
    \bm c= \frac{1}{\sqrt{3}}\begin{pmatrix}
        1 \\ 1 \\ 1 \\ 0
    \end{pmatrix}\,,
\end{align}
where $\bm a$ and $\bm b$ span the cross section.
The unit vector $\bm c$ represents the four-component generalization of the [111] vector orthogonal to the plane of the antiSkyrmion.
Then, we cut the antiSkyrmion along the $[-1,1,0]$ plane, as displayed in Fig.~\ref{fig:antiskyrmion-components}(c). This region shows $S^2$ Skyrmions in the effective moment $\tilde{\bm n}=(n_a,n_c,n_4)$.
A similar situation arises by cutting the antiSkyrmion along the [1,1,-2] plane, as reported in Fig.~\ref{fig:antiskyrmion-components}(d). In this case, we see $S^2$ Skyrmions for the effective moments $\tilde{\bm n}=(n_b,n_c,n_4)$.
Thus, the antiSkyrmion torus is simplified in three different $S^2$-like situations~\cite{gudnason15}.

Toroidal solitons in the context of the Skyrme-like model were already studied by one of the authors of this paper~\cite{gudnason15}. However, to the best of our knowledge, the antiSkyrmion found in the lattice formulation represents the first instance of a stable toroidal solution with the smallest nonzero topological charge $Q_{S^3}=-1$, which is in contrast to the usual case where toroidal solitons have topological charge $|Q_{S^3}|=2$ (this is the case for both Skyrmions and monopoles \cite{MantonSutcliffe}).

\section{Conclusion}
\label{sec:conclusion}

We have studied a class of sigma models from $\mathbb{R}^3$ to $S^3$
that generalize the theory of magnetic Skyrmions to three spatial
dimensions. The order parameter is a four-component unit vector
$\bm{n} \in S^3$, and the role of the DMI is played by one-derivative
terms that are invariant under the diagonal locked rotation group
$\SO(3)_{\rm diag}$. A systematic classification yields exactly two
inequivalent such terms, which have been denoted as the $\alpha$- and
$\beta$-terms~\cite{Gudnason:2024opf}. For each model we derived a
cubic-lattice discretization that reproduces the continuum theory at
long wavelengths, and we used Monte Carlo simulations to map its
finite-temperature phase diagram. The resulting phase structure is
substantially richer than that of the conventional $S^2$ magnetic
Skyrmion.

A feature shared by both models is the emergence of magnetic
strings. These are \emph{not} exact topological solitons -- since
$\pi_2(S^3)$ is trivial -- but they are dynamically stable over a
significant region of the phase diagram. Their stability originates
from a mechanism of \emph{restricted topology}: the negative
energy density of the string selects an equatorial $S^2\subset S^3$
as the preferred manifold, and the nontrivial $\pi_2(S^2)=\mathbb{Z}$
of this subspace supports a winding number that protects the
string against decay. The string tension is negative, with a
striking consequence for the exact topological charge of
$\pi_3(S^3)=\mathbb{Z}$: once a string is nucleated, it splits a
Skyrmion into two half-Skyrmions and expels them toward the boundary
of the system. We call this phenomenon \emph{anti-confinement}, by
analogy with -- but opposite to -- the confining string of QCD. Within
the string-lattice phase of the $\alpha$-model, the strings are free
to bend, forming zigzag patterns. String bends carry a fractional
$S^3$ topological charge; a simple geometric argument based on the
bending angle of approximately $120^\circ$ predicts $Q_{S^3}=\pm1/6$, in
good agreement with the lattice measurement $|Q_{S^3}|\approx0.19$.

For the $\alpha$-term model there exists, in addition, a pocket of
the phase diagram in which the Zeeman field is strong enough to
suppress string formation and stabilize a genuine three-dimensional
Skyrmion lattice. The Skyrmions are spherically symmetric, arrange
into a body-centered-cubic structure, and carry unit $\pi_3(S^3)$
charge, confirming the proposal of
Refs.~\cite{menta2023magnetic,Gudnason:2024opf}. As the Zeeman field
decreases (equivalently, as the effective DMI strength
$\tilde{\kappa}\propto B^{-1/2}$ increases), the Skyrmion lattice
gives way to the string-lattice phase through the anti-confinement
instability.

At still weaker fields, between the string-lattice and spin-spiral
phases, the $\alpha$-model hosts an antiSkyrmion lattice. Unlike the
spherical Skyrmion, the antiSkyrmion is toroidally shaped and retains
only axial symmetry, carrying negative unit $\pi_3(S^3)$ charge.
This lower symmetry is a direct consequence of the chirality of the
theory, which forces Skyrmions and antiSkyrmions to be inequivalent.
Its existence in the lattice phase diagram has no counterpart yet in
the continuum theory -- where the antiSkyrmion Ansatz has not been
analyzed -- and represents a concrete open problem that illustrates
the productive interplay between lattice simulation and analytical
methods.
To the best of our knowledge, this is the first instance of a stable toroidal soliton with (negative) unit charge.

Concerning physical realizations, the $S^3$ order-parameter manifold
arises naturally wherever the local degree of freedom is a rigid
four-component rotor, including spin-nematic and biaxial-nematic
phases of frustrated magnets and spinor condensates. A particularly
concrete route is offered by cold-atom platforms with synthetic
internal dimensions%
where a spinor Bose-Einstein condensate of $^{87}$Rb loaded into a
three-dimensional optical lattice, with Raman-assisted state-changing
tunneling, realizes the $\alpha$-term Hamiltonian exactly at the
mean-field level, with all three couplings $J$, $D$, and $B$
independently tunable. 

The present work establishes a theoretical and computational
framework for three-dimensional topological magnetic textures in
systems with $S^3$ order-parameter manifold. Several directions remain
open: a continuum analysis of the antiSkyrmion, a study of quantum
fluctuations around the mean-field cold-atom proposal, an
investigation of the half-Skyrmion bound states and their
braiding statistics, and the extension of the classification to
target spaces beyond $S^3$.

\begin{acknowledgments}
NF and RM thank the Galileo Galilei Institute (GGI) for Theoretical Physics for the hospitality during the “SFT 2024 - Lectures on Statistical Field Theories'' school and the INFN for partial support during the completion of this work.
We thank B.~Barton-Singer and L.~Janssen for insightful comments and collaboration on related projects.
This work has been supported by the Deutsche Forschungsgemeinschaft (DFG) through Project No.\ 247310070 (SFB 1143, A07), Project No.\ 390858490 (W\"urzburg-Dresden Cluster of Excellence \textit{ctd.qmat}, EXC 2147).
The work of SB is supported by the INFN special research project
grant ``GAST'' (Gauge and String Theories).
SBG thanks the Outstanding Talent Program of Henan University for partial support.
The authors gratefully acknowledge the computing time made available to them on the high-performance computer Barnard at the NHR Center of TU Dresden. This center is jointly supported by the Federal Ministry of Education and Research and the state governments participating in the National High-Performance Computing (NHR) joint funding program~\cite{nhr-alliance}.
\end{acknowledgments}
\bibliography{biblio}
\appendix
\section{Luttinger-Tisza method}
\label{appendix:LT-method}
The Luttinger-Tisza method is a powerful tool for studying the ground-state configurations of classical bilinear spin systems with periodic boundary conditions~\cite{luttinger46,lyons60,litvin74}. Consider a generic bilinear Hamiltonian
\begin{equation}
    \label{eq:generic-bil-ham}
    \mathcal{H}=\frac{1}{2}\sum_{i,j}\sum_{\mu,\nu}\bm{n}_{i,\mu}  \mathsf{J}_{i,\mu,j,\nu} \bm{n}_{j,\nu}\,,
\end{equation}
where the indices $i$ and $j$ run over the lattice unit cells, $i=1,\dots,\mathcal{V}$, while $\mu$ and $\nu$ are sublattice indices, $\mu=1,\dots,p$. The total number of sites is $N=p\mathcal{V}$, and each site has a position vector $\bm{r}_{i\mu}=\bm{R}_i+\boldsymbol{\delta}_\mu$.
The vector $\bm{R}_i$ specifies the position of the unit cell and can be expressed as a sum of primary vectors $\bm{t}_a$ with integer coefficients
\begin{equation}
    \label{eq:unit-cell-lattice-vec}
    \bm{R}_{i}=\sum_{a=1}^{d}m_i^{(a)}\bm{t}_a \quad\mathrm{with}\quad m_i^{(a)}\in\mathbb{Z}\,.
\end{equation}
Each spin on the lattice respects the strong spin constraint
\begin{equation}
    \label{eq:strong-spin}
    \bm{n}_{i,\mu}^2 = 1 \quad \forall\, i,\mu\,.
\end{equation}
Due to translational symmetry, the interaction matrix depends only on the relative position between two sites,
\begin{equation}
    \label{eq:interaction-mat-translation-prop}
    \mathsf{J}_{i,\mu,j,\nu}=\mathsf{J}_{\mu,\nu}(\bm{R}_i -\bm{R}_j)\,.
\end{equation}
The backbone of the method consists in the energy minimization of the Hamiltonian in Eq.~\eqref{eq:generic-bil-ham} with a modified spin constraint, namely the weak spin constraint,
\begin{equation}
    \label{eq:weak-spin}
    \sum_{i,\mu} \bm{n}_{i\mu}^2 = p\mathcal{V}=N\,,
\end{equation}
representing an averaged unit length of the spin squares.
By a Fourier transformation of the spin moments,
\begin{equation}
    \label{eq:spin-fourier}
    n_{j,\nu}^\alpha=\frac{1}{\sqrt{\mathcal{V}}}\sum_{j}e^{j\bm{k}\cdot \bm{R}_{j,\nu}}n_{\bm{k},\nu}^{\alpha}\,,
\end{equation}
the Hamiltonian is rewritten as a sum over momenta of the first Brillouin zone
\begin{equation}
    \label{eq:gen-bil-ham-momentum}
    \mathcal{H}=\frac{1}{2}\sum_{\bm{k}}\sum_{\mu,\nu}\bm{n}_{\bm{k},\mu}\mathsf{\Lambda}_{\mu,\nu}(\bm{k})\bm{n}_{-\bm{k},\nu}\,,
\end{equation}
with the interaction matrix in momentum space being 
\begin{equation}
    \label{eq:int-matrix-momentum}
    \mathsf{\Lambda}_{\mu,\nu}(\bm{k})=\sum_{\bm{r}}\mathsf{J}_{\mu,\nu}(\bm{r})e^{i\bm{k}\cdot (\bm{r}+\boldsymbol{\delta}_\mu-\boldsymbol{\delta}_\nu)}\,.
\end{equation}
Introducing the vector $\bm{Q}_{\bm{k}}=(\bm{n}_{\bm{k},1},\dots,\bm{n}_{\bm{k},p})$, the Hamiltonian acquires a diagonal form in momentum space
\begin{equation}
    \label{eq:LT-hamiltonian}
    \mathcal{H}=\frac{1}{2}\sum_{\bm{k}} \bm{Q}_{\bm{k}} \mathsf{\Lambda}(\bm{k})\bm{Q}_{-\bm{k}}\,.
\end{equation}
The interaction matrix $\mathsf{\Lambda}(\bm{k})$ in Eq.~\eqref{eq:LT-hamiltonian} is Hermitian, so it can be diagonalized by a unitary transformation, which respects the weak spin constraint in Eq.~\eqref{eq:weak-spin}. Considering spins of an arbitrary number of components $G$, there is a total of $pG$ eigenvalues $\lambda_{\mu}^{\alpha}$. 
The ground state is obtained by taking the minimal eigenvalue $\lambda_{\mathrm{min}}$ for some $\bm{k}_{\mathrm{min}}$, $\mu_{\mathrm{min}}$, and $\alpha_{\mathrm{min}}$. Once the minimal eigenvalue is found, the corresponding spin structure is obtained from the related eigenvector. Finally, the spin configuration is considered a physical ground state with ground state energy $E=pN\lambda_{\mathrm{min}}/2$ if it respects also the strong spin constraint in Eq.~\eqref{eq:strong-spin}.
%
%
%
%
\subsection{Application to the axial case with the \texorpdfstring{$\beta$-term}{beta-term}}
\label{subapp:axialDMI-LT}
The Hamiltonian in Eq.~\eqref{eq:axial-lattice-model} of Sec.~\ref{subsec:axial-DM-numerics},  reduces to a bilinear Hamiltonian of the kind as in Eq.~\eqref{eq:generic-bil-ham} if the Zeeman field is zero. The cubic lattice is a three-dimensional Bravais lattice, so $d=3$ and $p=1$, and we can drop the sublattice indices. The interaction matrix in momentum space is given by 
\begin{equation}
    \label{eq:axial-int-matrix}
    \begin{split}
        \mathsf{\Lambda}^{\alpha\beta}(\bm{k})=&-2J\delta^{\alpha\beta}\sum_{a=1}^{3}\cos{\bm{k}\cdot \hat{\bm{e}}_a} \\
        &+2iD\varepsilon^{\alpha\beta\rho\sigma}\sum_{\rho,\sigma}\Gamma_{\rho}\sum_{a=1}^{3}(\hat{\bm{e}}_a)_\sigma \sin{\bm{k}\cdot \hat{\bm{e}}_a}\,,
    \end{split}
\end{equation}
where the indices $\alpha,\beta,\rho,\sigma$ run over spin component. In this case, we consider four component spins, so $G=4$, and we expect four different eigenvalues. As in Sec.~\ref{subsec:axial-DM-numerics}, we take $\Gamma=(0,0,1,0)$. 
The matrix diagonalization returns four eigenvalue bands in momentum space. The minimization of the bands leads to conditions on the ordering wavevector $\bm{K}$
\begin{equation}
    \label{eq:LT-WV-conditions-axial}
    \begin{split}
        &\cos{K_x}=\cos{K_y}=\frac{1}{\sqrt{1+\kappa^2/2}}\,, \\
        &K_z=0\,,
    \end{split}
\end{equation}
where we introduced the ratio $\kappa=D/J$.
The corresponding ground state energy is $E=-JN(1+\sqrt{\kappa^2+4})$, and the normalized eigenvector corresponding to the minimal eigenvalue is given by
\begin{equation}
    \boldsymbol{\Psi}=\begin{pmatrix}
        i/2  \\ -i/2 \\  0  \\ 1/{\sqrt{2}}
    \end{pmatrix}\,.
\end{equation}
Hence, the spin configuration in direct space is described by a spin-spiral with wave vector $\bm{K}$,
\begin{eqnarray}
    \label{eq:axial-spin-spiral}
     \bm{n}_{i}&=&\frac{1}{\sqrt{2}}\left( \boldsymbol{\Psi}e^{i(\bm{K}\cdot \bm{R}_i + \varphi)} + \boldsymbol{\Psi}^* e^{-i(\bm{K}\cdot \bm{R}_i + \varphi)}  \right) \\ &=& \begin{pmatrix}
         -\sin({\bm{K}\cdot R_i + \varphi})/{\sqrt{2}} \\
         +\sin({\bm{K}\cdot R_i + \varphi})/{\sqrt{2}} \\
         0 \\
         \cos{(\bm{K}\cdot R_i + \varphi)}\\
     \end{pmatrix} \nonumber\,,
\end{eqnarray}
which respects the strong spin constraint in Eq.~\eqref{eq:strong-spin}. The phase $\varphi$ represents a free global phase: a change in the global phase returns a different spiral with the same energy and same wavevector $\bm{K}$. In conclusion, the spin-spiral is the ground state of the $\beta$-term model in Eq.~\eqref{eq:axial-lattice-model} with $\bm{B}=0$ and periodic boundary conditions.
With the set of parameters used in \ref{subsec:axial-DM-numerics}, $\Gamma=(0,0,1,0)$, $J=1$, and $\kappa=\sqrt{2}$ the spin-spiral has an ordering wavevector $\bm{K}=(\pm \pi/4, \pm\pi/4,0)$.

The spin-spiral state is topologically trivial, with the $\pi_3(S^3)$ topological charge in Eq.~\eqref{eq:S3-topological-charge} being exactly zero.
Furthermore, it has also zero $\pi_2(S^2)$ topological charge as defined in Eq.~\eqref{eq:S2-topological-charge}.
%
%
%
%
\subsection{Application to the spherical case with the \texorpdfstring{$\alpha$}{alpha}-term}
\label{subapp:spherical-DMI-LT}
The Hamiltonian in Eq.~\eqref{eq:spherical-lattice-model} reduces to a bilinear model as in Eq.~\eqref{eq:generic-bil-ham} if the Zeeman term vanishes. In the numerical calculations, we considered a three-dimensional cubic lattice, so $d=3$ and $p=1$, and we can drop the sublattice index also in this case. The $4\times4$ interaction matrix reads
\begin{align}
            \label{eq:spherical-int-mat}
        \mathsf{\Lambda}^{\alpha\beta}(\bm{k})=&-2J\delta^{\alpha\beta}\sum_{a=1}^{3}\cos{\bm{k}\cdot \hat{\bm{e}}_a} \\
        &+2iD\sum_{a=1}^{3}\left[(\hat{\bm{e}}_a)^\alpha (\hat{\bm{e}}_4)^\beta - (\hat{\bm{e}}_4)^\alpha (\hat{\bm{e}}_a)^\beta \right] \sin{\bm{k}\cdot \hat{\bm{e}}_a}\,,\nonumber
\end{align}
where $\alpha$ and $\beta$ are spin-component indices. The minimization of the corresponding eigenvalues translates to a condition on the ordering wavevector $\bm{K}$,
\begin{equation}
    \label{eq:LT-WV-conditions-spherical}
    \cos{K_x}=\cos{K_y}=\cos{K_z}=\frac{1}{\sqrt{1+\kappa^2/3}}\,,
\end{equation}
with a ground state energy of $E=-JN\sqrt{3(\kappa^2+3)}$. The normalized eigenvector related to the minimal eigenvalue is
\begin{equation}
    \label{eq:LT-spherical-eigenvector}
    \boldsymbol{\Psi}=\frac{1}{\sqrt{2}}\begin{pmatrix}
        i/\sqrt{3} \\ i/\sqrt{3} \\ i/\sqrt{3} \\ 1
    \end{pmatrix}
\end{equation}
so the ground state is described by a spin spiral with wavevector $\bm{K}$ and spin configuration
\begin{eqnarray}
    \label{eq:spherical-spin-spiral}
    \bm{n}_{i}&=&\frac{1}{\sqrt{2}}\left( \boldsymbol{\Psi}e^{i(\bm{K}\cdot \bm{R}_i + \varphi)} + \boldsymbol{\Psi}^* e^{-i(\bm{K}\cdot \bm{R}_i + \varphi)}  \right) \\ &=& \begin{pmatrix}
         -\sin({\bm{K}\cdot R_i + \varphi})/{\sqrt{3}} \\
         -\sin({\bm{K}\cdot R_i + \varphi})/{\sqrt{3}} \\
         -\sin({\bm{K}\cdot R_i + \varphi})/{\sqrt{3}} \\
         \cos{(\bm{K}\cdot R_i + \varphi)}\\
     \end{pmatrix}\,\nonumber,
\end{eqnarray}
where the phase $\varphi$ is the same free phase introduced in Eq.~\eqref{eq:axial-spin-spiral}.
In the main text, we used the set of parameters $J=1$ and $\kappa=\sqrt{3}$, so the spin spiral in the $\bm{B}=0$ limit with periodic boundary conditions propagates with a wavevectors $\bm{K}=(\pm \pi/4,\pm \pi/4, \pm\pi/4)$.
The $\pi_3(S^3)$ topological charge and the $\pi_2(S^2)$ topological charge are exactly zero on the spin-spiral described in Eq.~\eqref{eq:spherical-spin-spiral}.

\section{Spin spiral in a field: continuum calculation with spatial approximation}
\label{appendix:spin-spiral-continuum-method}
In this Appendix, we cover the steps used to estimate the spin-spiral energy in the presence of an external Zeeman field. This estimate allows a comparison between the spin-spiral state and the string lattice phase in Sec.~\ref{subsec:axial-DM-numerics}, helping in the identification of the energy-favorable phase in a phase-competing scenario.

Assuming that the spin spiral realized at $\bm{B}=0$ survives at small finite fields, it is possible to write a generalized spin spiral in the continuum~\cite{han2017Skyrmions},
\begin{equation}
    \label{eq:spin-spiral-continuum}
    \bm{n}(\bm{r})= \bm{n}_0 + \boldsymbol{\Psi}e^{i(\bm{K}\cdot\bm{r} + \varphi)} + \boldsymbol{\Psi}^*e^{-i(\bm{K}\cdot\bm{r} + \varphi)}\,,
\end{equation}
where the uniform magnetization $\bm{n}_0 \parallel \bm{B}$ is due to canting of the spins towards the external field caused by the Zeeman term. Here, we consider a spiral is assumed with single wave vector $\bm{K}$. 
The magnetic moments are subject to the normalization condition 
\begin{equation}
    \label{eq:continuum-normalization}
    \begin{split}
        \bm{n}^2&=\bm{n}_0^2 + 2\boldsymbol{\Psi}\cdot\boldsymbol{\Psi}^* \\
        &+ \boldsymbol{\Psi}\cdot\boldsymbol{\Psi}e^{2i(\bm{K}\cdot\bm{r}+\varphi)}+\boldsymbol{\Psi}^*\cdot\boldsymbol{\Psi}^*e^{-2i(\bm{K}\cdot\bm{r}+\varphi)} \\
        &+ 2\bm{n}_0\cdot\boldsymbol{\Psi}e^{i(\bm{K}\cdot\bm{r}+\varphi)}+ 2\bm{n}_0\cdot\boldsymbol{\Psi}^*e^{-i(\bm{K}\cdot\bm{r}+\varphi)}=1\,.
    \end{split}
\end{equation}
The energy density $\mathcal{E}$ consists of three terms given by the Heisenberg ($\mathcal{E_2}$), Dzyaloshinskii–Moriya ($\mathcal{E}_1$), and Zeeman ($\mathcal{E}_0$) contributions
\begin{equation}
    \mathcal{E}=\mathcal{E}_{2}+\mathcal{E}_{1}+\mathcal{E}_{0}\,,
\end{equation}
and it can be evaluate on the spin-spiral Ansatz in Eq.~\eqref{eq:spin-spiral-continuum}. In our case, the Heisenberg and Zeeman term are
\begin{equation}
    \label{eq:continuul-common-ene-dens-terms}
    \begin{split}
        \mathcal{E}_2 &=\frac{Ja^{-1}}{2}\sum_{\mu}\partial_\mu \bm{n} \cdot \partial_\mu \bm{n}\,, \\
        \mathcal{E}_0 &= Ba^{-3}(1-\bm{n}\cdot\bm{N})\,,
    \end{split}
\end{equation}
and they remain the same in both the $\alpha$ and $\beta$ terms for the DM interaction.

In general, the problem requires the introduction of a Lagrange multiplier $\lambda$ in the generalized energy density $\mathcal{L}$,
\begin{equation}
    \label{eq:continuum-generalized-ene-dens}
    \mathcal{L}=\mathcal{E}+\lambda(\bm{n}^2-1)
\end{equation}
to account for the normalization constraint on the magnetic moments, with the additional condition $\partial\mathcal{L}/\partial\lambda=0$. 
Then, a minimization procedure leads to the spin spiral parameters $\bm{n_0}$, $\boldsymbol{\Psi}$, $\bm{K}$, and the Lagrange multiplier $\lambda$  minimizing the energy density.
Typically, the problem cannot be easily solved if $B\neq0$. However, the condition in Eq.~\eqref{eq:continuum-normalization} introduces many oscillating factors. Since the Ansatz is periodic, the oscillating terms vanishes once we average the spiral on its unit cell. This suggests to tackle the minimization problem in Eq.~\eqref{eq:continuum-generalized-ene-dens} within a spatial-average approximation, with a generalized energy density
\begin{equation}
    \label{eq:continuum-averaged-ene-dens}
    \overline{\mathcal{L}} = \overline{\mathcal{E}} + \lambda(\overline{\bm{n}^2}-1)\,,
\end{equation}
where the overline represents the averaged quantities, and the averaged normalization condition in Eq.~\eqref{eq:continuum-normalization} simplifies to 
\begin{equation}
    \label{eq:averaged-norm-condition}
    \overline{\bm{n}^2}=\bm{n}_0^2+2\boldsymbol{\Psi}\cdot\boldsymbol{\Psi}^*\,.
\end{equation}
In our approximation, we solved
\begin{equation}
    \label{eq:continuum-minimizing-eq}
    \frac{\partial\overline{\mathcal{L}}}{\partial n_0^\rho}=0\,, \quad \frac{\partial\overline{\mathcal{L}}}{\partial \psi^\alpha}=0\,,\quad\frac{\partial\overline{\mathcal{L}}}{\partial\lambda}=0\,,
\end{equation}
 to express $\bm{n}_0$, $\boldsymbol{\Psi}$, and $\lambda$ as functions of the unknown wavevector $\bm{K}$. 
 Then, we minimized the generalized energy density in Eq.~\eqref{eq:continuum-averaged-ene-dens} with respect to $\bm{K}$, to get the ordering wavevector of the spin spiral.
At the end of the procedure, we have the spin-spiral parameters $\bm{n}_0$, $\boldsymbol{\Psi}$, the ordering wavevector $\bm{K}$, and the Lagrange multiplier $\lambda$, allowing the estimate of the averaged energy density $\overline{\mathcal{E}}$.
%
%
\subsection{Application to the axial case with the \texorpdfstring{$\beta$}{beta}-term}
The model with the $\beta$-term has a DM energy density
\begin{equation}
    \label{eq:continuum-axial-DM-ene-dens}
    \mathcal{E}_{1}=Da^{-2}\sum_{\alpha\beta\sigma\rho}\epsilon^{\alpha\beta\sigma\rho}\Gamma_\alpha n_\beta \partial_\sigma n_\rho \,,
\end{equation}
where we considered the vector $\bm{N}=(0,0,0,1)$ and $\boldsymbol{\Gamma}=(0,0,1,0)$.
Solving Eq.~\eqref{eq:continuum-minimizing-eq} and minimizing with respect to the wavevector $\bm{K}$ leads to 
\begin{equation}
    \label{eq:continuum-axial-spin-spiral}
    \begin{split}
        \bm{n}_0 &= \frac{B}{\kappa D}\bm{N}\,, \qquad
        \boldsymbol{\Psi} = \frac{1}{2}\sqrt{1-\left(\frac{B}{\kappa D} \right)^2} \begin{pmatrix}
            -i\sin{\theta}  \\ i\cos{\theta} \\ 0 \\ 1
        \end{pmatrix}\,, \\
        \bm{K} &=\kappa\begin{pmatrix}
            \cos{\theta} \\
            \sin{\theta} \\
            0
        \end{pmatrix}\,, \quad\,
        \lambda = \frac{D^2 a^{-3}}{2J}\,, \\
    \end{split}
\end{equation}
where $\kappa=D/J$. The energy density is given by 
\begin{equation}
    \label{eq:continuum-axial-ene-dens-result}
    \mathcal{E}=-\frac{Ja^{-3}}{2}\left(\kappa -\frac{B}{D} \right)^2\,.
\end{equation}
The emergent U(1) symmetry parametrized by the angle $\theta$ is due to the spin-orbit axial symmetry of the DM term in Eq.~\eqref{eq:continuum-axial-DM-ene-dens} \cite{Gudnason:2024opf,menta2023magnetic}, and it is broken once lattice anisotropies are taken into account, for example, with a fourth-order term in the magnetic moments~\cite{han2017Skyrmions}.

The estimate of the energy density in Eq.~\eqref{eq:continuum-axial-ene-dens-result} works for small Zeeman fields. In particular, from Eq.~\eqref{eq:continuum-axial-spin-spiral}, we have $\boldsymbol{\Psi}\cdot\boldsymbol{\Psi}=0$, so the approximation effectively neglects only the terms proportional to $2\bm{n}_0\cdot \boldsymbol{\Psi}$, which are of order 
\begin{equation*}
    4\frac{B}{\kappa D}\frac{1}{2}\sqrt{1-\left(\frac{B}{\kappa D}\right)^2}\approx \frac{2B}{\kappa D}
\end{equation*}
so the estimate is valid for $B\ll\kappa D/2$.
%
%
\subsection{Application to the spherical case with the \texorpdfstring{$\alpha$}{beta}-term}
In the $\alpha$-term model, the DM contribution reads
\begin{equation}
    \label{eq:continuum-spherical-DM-ene-dens}
    \mathcal{E}_{1}=Da^{-2}\sum_{\mu}\left(n^4 \partial_\mu n^\mu -n^\mu\partial_\mu n^4 \right) \,.
\end{equation}
The solution of Eq.~\eqref{eq:continuum-minimizing-eq} together with the minimization with respect to the wavevector $\bm{K}$ results in a spin spiral
\begin{align}
        \bm{n}_0 &= \frac{B}{\kappa D}\bm{N}\,, \qquad
        \boldsymbol{\Psi} = \frac{1}{2}\sqrt{1-\left(\frac{B}{\kappa D} \right)^2} \begin{pmatrix}
            i\sin{\theta}\cos{\phi}  \\ 
            i\sin{\theta}\sin{\phi} \\
            i\cos{\theta} \\
             1
        \end{pmatrix}\,, \non
        \bm{K} &=\kappa\begin{pmatrix}
            \sin{\theta}\cos{\phi}  \\ 
            \sin{\theta}\sin{\phi} \\
            \cos{\theta} \\
        \end{pmatrix}\,, \quad
        \lambda = \frac{D^2 a^{-3}}{2J}\,, 
            \label{eq:continuum-spherical-spin-spiral}
\end{align}
where the angles $(\theta,\phi)$ parameterize the O(3) symmetry due to the form of the residual spin-orbit coupling in Eq.~\eqref{eq:continuum-spherical-DM-ene-dens} \cite{han2017Skyrmions,menta2023magnetic,Gudnason:2024opf}. With these parameters, the energy density is given by
\begin{equation}
    \label{eq:continuum-spherical-ene-dens-result}
    \varepsilon=-\frac{Ja^{-3}}{2}\left(\kappa -\frac{B}{D} \right)^2\,.
\end{equation}
Again, the approximation is valid for $2B/\kappa D\ll1$.

Interestingly, besides the angle structure, the results are identical to the axial case. This might have to do with the simple, highly symmetric cases considered here. The most generic DM tensor in Eq.~\eqref{eq:Theta_inv_standard} might lead to different results.
%
\section{Alignment between SO(3) subgroup of magnetization vector and string direction}
\label{appendix:alignment_strings}

Let us consider the Ansatz for a string in the continuum in the case of the spherical DMI in Eq.~\eqref{eq:DMI_sph}:
\begin{eqnarray}
\bn = \begin{pmatrix}
\sin\vartheta\cos\phi\\
\sin\vartheta\sin\phi\\
0\\
\cos\vartheta
\end{pmatrix},
\end{eqnarray}
where $\vartheta$ is the string profile function in the string plane that has coordinate $r$ and $\phi$ and whose orientation in $\mathbb{R}^3$ are yet to be determined.
More precisely, the string plane is given by the coordinate
\begin{align}
r &= \sqrt{(x')^2 + (y')^2},\qquad
\phi = \arctan(y'/x'),
\end{align}
and the primed coordinates are related to the standard Cartesian coordinates by two rotation matrices
\begin{eqnarray}
\begin{pmatrix}
x'\\
y'\\
z'
\end{pmatrix}
=
R_z(\beta) R_x(\alpha)
\begin{pmatrix}
x\\
y\\
z
\end{pmatrix},
\end{eqnarray}
where the rotation SO(3) matrices are given by
\begin{align}
R_x(\alpha) &= 
\begin{pmatrix}
1 & 0 & 0\\
0 & \cos\alpha & -\sin\alpha\\
0 & \sin\alpha & \cos\alpha
\end{pmatrix}, \quad
R_z(\beta) = 
\begin{pmatrix}
\cos\beta & -\sin\beta & 0\\
\sin\beta & \cos\beta & 0\\
0 & 0 & 1
\end{pmatrix}.
\end{align}
These two angles are sufficient for the most general rotation of a plane, since we do not need a rotation of the plane about its perpendicular axis here.
Denoting the Cartesian partial derivatives as $\p_i=\frac{\p}{\p x^i}$, we can easily verify that
\begin{align}
(\p_i x')^2 = 
(\p_i y')^2 = 
(\p_i r)^2  = 1,\ \ 
(\p_i \phi)^2 = \frac{1}{r^2},\ \ 
\p_i x'\p_i y' = 0.\nonumber
\end{align}
Hence, we get that the kinetic (Heisenberg exchange) term is simply
\begin{eqnarray}
\frac12(\p_i\bn)^2 = \frac12(\vartheta'(r))^2 + \frac{1}{2r^2}\sin^2(\vartheta),
\end{eqnarray}
and is therefore independent of the rotation angles of the plane, $\alpha$ and $\beta$.

The potential term is unaffected by the rotation of the plane too, since it is a function of $\cos\vartheta$ only.

Any dependence of the string-plane's orientation must hence reside within the DMI.
A straightforward computation yields
\begin{align}
&\kappa(n^4\nabla\cdot\bn - \bn\cdot\nabla n^4)\non
&= \kappa\Big(
\cos\phi\vartheta'(r)\p_1 r + \sin\phi\vartheta'(r)\p_2 r\non
&\phantom{=\kappa\Big(}
+\frac12\cos\phi\sin(2\vartheta)\p_2\phi
-\frac12\sin\phi\sin(2\vartheta)\p_1\phi
\Big)\non
&=\kappa\Big(
\vartheta'(r)\Big[
\cos^2\phi\p_1x'+\sin^2\phi\p_2y'+\frac{\sin(2\phi)}{2}(\p_1y'+\p_2x')
\Big]\non
&\phantom{=\kappa\Big(}
+\frac{\sin(2\vartheta)}{2r}\Big[
\cos^2\phi\p_2y'+\sin^2\phi\p_1x'
-\frac{\sin(2\phi)}{2}(\p_2x'+\p_1y')
\Big]
\Big)\non
&=\kappa\Big(
\vartheta'(r)\big[
\cos\beta\cos^2\phi+\sin^2\phi
\big]\non
&\phantom{=\kappa\Big(}
+\frac{\sin(2\vartheta)}{2r}\big[
\cos\alpha\cos\beta\cos^2\phi + \sin^2\phi
\big]
\Big).
\end{align}

The DMI energy is negative due to $\vartheta'(r)<0$, so the maximally negative DMI energy corresponds to $\beta=0$.
Similarly, one can deduce from the fact that $\sin(2\vartheta)<0$ in the core of the string where $r$ is small and is positive only where $r$ is large, so this term too corresponds to maximally negative DMI energy when $\alpha=\beta=0$.
This corresponds simply to $(x',y',z')=(x,y,z)$ and hence the string is not rotated and lies in the $(x,y)$ plane. Notice this orientation is fixed by the Ansatz where $n^3=0$.
By permutation of the fields, one obtains the alignment where the preferred string-plane is given by the plane orthogonal to the vanishing component of the magnetization vector $n^i$.

\end{document}

%% file: commands.tex
\newcommand{\beq}{\begin{eqnarray}}
\newcommand{\eeq}{\end{eqnarray}}
\newcommand{\non}{\nonumber\\}

\DeclareMathOperator{\Og}{O}
\DeclareMathOperator{\SO}{SO}

\newcommand{\p}{\partial}

\renewcommand{\d}{\mathop{}\!\mathrm{d}}

\newcommand{\bn}{\mathbf{n}}

\newcommand{\bea}{\begin{eqnarray}}
\newcommand{\eea}{\end{eqnarray}}
\newcommand{\be}{\begin{equation}}
\newcommand{\ee}{\end{equation}}

\def\nn{\nonumber}

\def\E{{\cal E}}